\documentclass[prb,superscriptaddress,showpacs,longbibliography,reprint]{revtex4-2}

\usepackage{hyperref}

\usepackage{color}
\usepackage[usenames,dvipsnames]{xcolor}
\usepackage{amsmath,amsthm,amssymb}
\usepackage{graphicx}
\usepackage{epsfig}
\usepackage{dcolumn}
\usepackage{bm}
\usepackage{mathrsfs}
\usepackage{multirow}
\usepackage[all]{xy}
\usepackage{pbox}
\usepackage{verbatim}
\usepackage{braket}
\usepackage{dsfont}

\DeclareMathOperator{\Tr}{Tr}

\usepackage{graphicx}
\graphicspath{{./Figures/}}

\usepackage{color,soul}

\begin{document}

\title{Dissipative quasi-particle picture for quadratic Markovian open quantum systems}
\author{Federico Carollo}
\affiliation{Institut f\"{u}r Theoretische Physik, Universit\"{a}t T\"{u}bingen, Auf der Morgenstelle 14, 72076 T\"{u}bingen, Germany}
\author{Vincenzo Alba}
\affiliation{Dipartimento di Fisica, Universit\`a di Pisa, and INFN Sezione di Pisa, Largo Bruno Pontecorvo 3, Pisa, Italy}

\begin{abstract}
Correlations between different regions of a quantum many-body system can be quantified through measures based on entropies of (reduced) subsystem states.  For closed systems, several analytical and numerical tools, e.g., hydrodynamic theories or tensor networks, can accurately capture the time-evolution of subsystem entropies, thus allowing for a profound understanding of the unitary dynamics of quantum correlations. However,  these methods either cannot be applied to open quantum systems or do not permit an efficient computation of quantum entropies for mixed states. Here, we make progress in solving this issue by developing a dissipative quasi-particle picture ---describing quantum entropies and the mutual information in the limit of large space-time coordinates with their ratio being fixed--- and showing its validity for quadratic open quantum systems. Our results demonstrate that the open quantum many-body dynamics of correlations can be understood in terms of propagating (dissipative) quasi-particles. \end{abstract}

\maketitle

\section{Introduction}
Entropy has  a fundamental role in science \cite{wehrl1978}. In thermodynamics, it provides the arrow of time while in information theory, it quantifies the uncertainty associated with a random variable \cite{shannon1948}. Entropic functionals capture correlations, e.g., through the mutual information \cite{cover1991,mezard2009,gray2011}, and can even characterize entanglement. This aspect is crucial for the study of nonequilibrium universal behavior in the spreading of correlations, either under unitary \cite{calabrese2005,chiara2006,calabrese2007,lauchli2008,manmana2009,cheneau2012,calabrese2012,calabrese2012b,jurcevic2014,richerme2014,islam2015,eisert2015,kaufman2016,chang2019,rakovszky2019,brydges2019,elben2020,gillman2021,denicola2021,alba2021b,mendozaarenas2021,neel2021}, dissipative \cite{bernier2018,macieszczak2019,malouf2020,maity2020,alba2021,rossini2021} or stochastic \cite{li2018,znidaric2020,chan2019,skinner2019,cao2019,jian2020,carollo2020,piroli2020,ippoliti2021,nahum2021,alberton2021,lavasani2021,bernard2021}
 dynamics.
 
\begin{figure}[t]
\centering
\includegraphics[width=\linewidth]{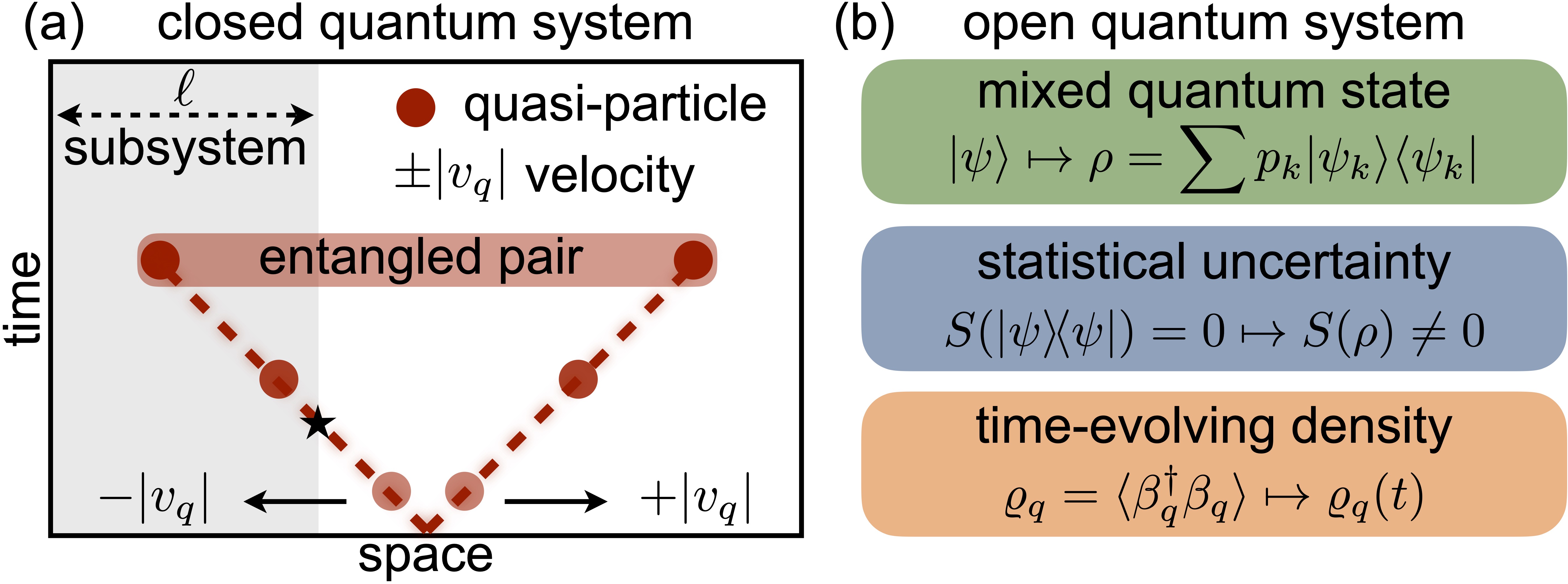}
\caption{{\bf Quasi-particle picture for closed systems and irreversible effects.} a)  A subsystem of length $\ell$ is embedded in a closed many-body system.  Quasi-particles of an initially entangled pair travel in opposite directions, with velocity $\pm |v_q|$. When one of them enters the subsystem (see black star), correlations of the pair contribute to the subsystem entropy.  b) For open systems, the quantum state is mixed and, thus, subsystem entropies are affected by statistical uncertainty. Furthermore, quasi-particle densities are not conserved [cf.~Eq.~\eqref{density}] but rather obey a rate equation.}
\label{Fig1}
\end{figure}

Here, we consider a dissipative nonequilibrium setup: a many-body  system undergoes an open quantum Lindblad dynamics \cite{lindblad1976}, combining coherent and irreversible effects, and is initialized in a state which is not a stationary state of the Hamiltonian nor of the full Lindblad dynamics. We focus on a subsystem embedded in this system, see Fig.~\ref{Fig1}(a), and we are concerned with the time-evolution of subsystem entropies and of the mutual information of the bipartition. For closed integrable systems,  in the limit of large space-time coordinates with their ratio being fixed, this is typically described by a quasi-particle picture \cite{calabrese2005,calabrese2007,fagotti2008,alba2017,alba2017b,alba2018,alba2019b}. 
The initial state acts as a source of entangled pairs of quasi-particle excitations  ---labelled by the quasi-momentum $q$. These propagate ballistically in opposite direction with velocity $\pm |v_q|$ [see Fig.~\ref{Fig1}(a)] and, when shared by the subsystem and the remainder of the many-body system, contribute to the subsystem entropy  through their correlation content \cite{castro-alvaredo2018}.
This picture (see Refs.~\cite{bertini2018,bertini2018b,bastianello2018,alba2019,bastianello2020} for extensions) has proved valuable in understanding the dynamics after quantum quenches as well as the approach to thermodynamic ensembles in closed systems \cite{alba2017}. 
However, its applicability to nonequilibrium systems undergoing dissipative dynamics is far from clear, and, thus, studying correlations in open quantum systems remains a challenging task. 

In this paper, we introduce a dissipative quasi-particle picture accounting for irreversible effects associated with open quantum Lindblad dynamics [cf.~Fig.~\ref{Fig1}(b)]. We show that it accurately predicts the time-evolution of subsystem entropies and, importantly, of correlations for quadratic open quantum systems,  within the setup discussed above, starting from nonstationary states with low correlations. The essence of our picture is encoded in the formula in Eq.~\eqref{formula}, which is a conjecture that we make based on the results derived in Ref.~\cite{alba2021} for a specific fermionic system. As we discuss in this paper, the  R\'enyi-$n$ (and von Neumann) entropy $S^{(n)}_\ell$ of a subsystem of length $\ell$, obeys, at leading order in $\ell$ and for times of order $t\propto \ell$ (see below), the relation 
\begin{equation}
\begin{split}
S^{(n)}_\ell(t)=&\int \frac{dq}{2\pi}\Big\{\ell s^{(n),\,  {\rm mix}}_q(t)+\\
+&\min(2|v_q|t,\ell)\!\left[s_q^{(n),\, {\rm YY}}(t)-s_q^{(n),\, {\rm mix}}(t)\right]\Big\}.
\end{split}
\label{formula}
\end{equation}
Here, $s_q^{(n),\, {\rm mix}},s_q^{(n),\, {\rm YY}}$ are two different entropic contributions. The first, $s_q^{(n),\, {\rm mix}}$, accounts for the mixedness of the $q$th quasi-particle state [cf.~Fig.~\ref{Fig1}(b)]. 
The quantity $s_q^{(n),\, {\rm YY}}$ [see Eq.~\eqref{s_quantum_q} below] is instead the quasi-particle contribution to the Yang-Yang entropy \cite{yang1969}, which is the entropy quantifying the number of microscopic states that, in the thermodynamic limit, give rise to a same macroscopic state specified by the quasi-particle densities. In closed systems,  this macroscopic state is the generalized Gibbs ensemble describing local stationary properties of the system \cite{polkovnikov2011,calabrese2016,essler2016,vidmar2016,caux2013,caux2016}, and quasi-particles are given by the Hamiltonian eigenmodes. In our setting, quasi-particles are instead defined by the eigenmodes of the Lindblad dynamics and their density is in general time-dependent [see Fig.~\ref{Fig1}(b)]. 

The formula in Eq.~\eqref{formula} has a transparent physical interpretation. The first term accounts for the contribution due to statistical uncertainty of the full system state [cf.~Fig.~\ref{Fig1}(b)]. The second term is related to the ballistic propagation of quasi-particles.  The ``$\min$" function counts the pairs  shared by the bipartition at time $t$ \cite{calabrese2005} [cf.~Fig.~\ref{Fig1}(a)], while the square brackets provide their correlation content. As for closed systems, the latter depends on $s_q^{(n),\, {\rm YY}}$ but it is here diminished by statistical uncertainty as quantified by $s_q^{(n),\, {\rm mix}}$. While, for mixed states, subsystem entropies do not measure correlations, our formula allows us to ``extract" from them the appropriate contribution uniquely associated with correlations: as we show, the second term in Eq.~\eqref{formula} [see Eq.~\eqref{propag}] is indeed equivalent to the mutual information between the subsystem and the remainder of the system. Our approach thus introduces a powerful method for describing  correlations in  open quantum many-body systems. 

Below, we provide Eq.~\eqref{formula} with predictive power, deriving all relevant terms for quadratic systems \cite{prosen2008,prosen2010b,prosen2010,kos2017,guo2017}. We focus the presentation on bosonic ones. 

\section{Quadratic dissipative dynamics} 
\label{sec_Lind} We consider translation invariant one-dimensional quantum systems made of $L$ sites. Each site is occupied by a bosonic mode, described by the operators $x_i,p_i$, such that $[x_i,p_j]={\rm i}\delta_{ij}$. We collect these operators in the column vector ${r}=(x_1,p_1,x_2,p_2,\dots x_L,p_L)$. The commutation relations are expressed as $[r_i,r_j]={\rm i}\Omega_{ij}$, where $\Omega$ is a block matrix with blocks given by 
$$
\lceil\Omega\rfloor_{ij}=\delta_{ij}\, \sigma \, ,\quad \mbox{ where } \quad \sigma=\begin{pmatrix}
0&1\\
-1&0
\end{pmatrix}\, .
$$
For a given matrix $M$ we denote matrix elements as $M_{ij}$, while we denote its $i,j$th $2\times2$ block as $\lceil M\rfloor_{ij}$. 

The many-body system undergoes a gaussian Markovian open quantum time-evolution \cite{demoen1979,holevo2001,hellmich2010,heinosaari2010,weedbrook2012,parthasarathy2015}. The dynamics of any operator $O$ is  implemented by the master equation $\dot{O}_t=\mathcal{L}[O_t]$, with Lindblad generator \cite{lindblad1976,gorini1976}
\begin{equation}
\mathcal{L}[O]=i[E,O]+\sum_{i,j=1}^{2L}C_{ij}\left(r_iOr_j-\frac{1}{2}\left\{O,r_ir_j\right\}\right)\, .
\label{Lindblad}
\end{equation}
The system Hamiltonian $E=\sum_{i,j=1}^{2L}H_{ij}\, r_ir_j$ is quadratic with $H=H^T$, and $T$ denotes transposition. The positive semi-definite matrix $C$ accounts for irreversible effects and decomposes as $C=A+{\rm i}B$, with $A$ being real symmetric and $B$ real anti-symmetric. 

Translation invariance requires $H$ and $C$ to be block-circulant, i.e.~matrices of the form 
\begin{equation}
M=\begin{pmatrix}
m_0&m_1&m_2&&\dots &m_{L-1}\\
m_{L-1}&m_0&m_1&m_2&&\vdots\\
&m_{L-1}&m_0&m_1&\ddots\\
\vdots & \ddots &\ddots & \ddots& &m_2\\
&&&&&m_1\\
m_1&\dots &&&m_{L-1}&m_0
\end{pmatrix}\, , 
\label{circ_M}
\end{equation}
with $m_i$ being $2\times2$ matrices. Such block-circulant matrices are thus fully specified by $L$, in principle different, $2\times2$ matrices, which provide all their blocks according to the relation
\begin{equation}
\lceil H\rfloor_{ij}=h_{{\rm mod}(j-i,L)} \quad \mbox{ and } \quad  \lceil C\rfloor_{ij}=c_{{\rm mod}(j-i,L)}\, .
\label{M}
\end{equation}
Each of these blocks describes how sites $i$ and $j$ are (either coherently or dissipatively) coupled. Any block-circulant matrix, like the matrix $M$ above, becomes block-diagonal under rotation with the Fourier-transform unitary operator $U$, with $\lceil U\rfloor_{kj}=e^{{\rm i} q_k j } {\bf 1}_2$, where ${\bf 1}_2$ is the $2\times2$ identity and $q_k=2\pi k/L$ the quasi-momenta. That is, $\hat{M}:=UMU^\dagger$ has only blocks on the diagonal --- so-called symbols --- given by
\begin{equation}
\lceil \hat{M}\rfloor_{kk}:=\hat{m}_{q_k}=\frac{1}{L}\sum_{i,j=1}^Le^{{\rm i}q_k(i-j)}m_{{\rm mod}(j-i,L)}\, .
\label{symbol}
\end{equation} 

\subsection{Examples of dissipation}
The map in Eq.~\eqref{Lindblad} can account for completely generic quadratic dissipative processes. For concreteness, we discuss here in details the form assumed by the generator for the most common dissipative processes. We consider pump and loss of bosonic excitations at rates $\gamma^+$ and $\gamma^-$, diffusion in momentum space at rate $\gamma^{x}$ (implemented through operators $x_i$) and in position space at rate $\gamma^p$ (implemented through operators  $p_i$). However, our approach is very general and is not restricted to these cases. 
We further allow for dissipation to be non-local, i.e.~not occurring independently from site to site. The matrix $C$ consists of the combination of different processes. For example, one may have $C=\sum_\alpha C^{\alpha}$ where $C^\alpha$, with $\alpha=+,-,x,p$, are associated with the processes mentioned above and read as $\lceil C^{\alpha}\rfloor_{ij}=\gamma^\alpha f^\alpha_{ij} c^\alpha$, with
\begin{equation}
c^\pm= \frac{1}{2}\begin{pmatrix}
1&\mp {\rm i}\\
\pm {\rm i}&1
\end{pmatrix} ,\quad c^x=\begin{pmatrix}
1&0\\
0&0
\end{pmatrix} ,\quad c^p=\begin{pmatrix}
0&0\\
0&1
\end{pmatrix} .
\label{com-proc}
\end{equation}
Here, the functions $f^\alpha_{ij}=f^\alpha(d_{ij})$ solely depend on the distance $d_{ij}=\min[|i-j|,L-|i-j|]$ between sites.

We now present an explicit calculation of the the symbols $\hat{a}_{q_k}$ and $\hat{b}_{q_k}$, for the above processes, assuming for concreteness an even number of sites $L=2n$. We first provide the blocks on the diagonal of the matrices $\hat{C}^\alpha$. These are given by 
$$
\lceil\hat{C}^\alpha\rfloor_{kk}:= \hat{c}_{q_k}^\alpha = c^\alpha \, \mathcal{S}_k[f^\alpha]\,  , 
$$
where $\mathcal{S}_k[f]$ is a (functional) quantity determined by the function $f$ and the parameter $k$ as 
\begin{equation}
\mathcal{S}_k[f^\alpha]=\left(f^\alpha(0)+(-1)^k f^\alpha(n)+2\sum_{j=1}^{n-1}f^\alpha(j)\cos (q_k j)\right)\, .
\label{eq-functional-sm}
\end{equation}
The Fourier transform of the symmetric part of the complete dissipative matrix $C=C^++C^-+C^x+C^p$, which is defined as $A=(C+C^T)/2$, is given by 
\begin{equation}
\begin{split}
\hat{a}_{q_k}&=\frac{\gamma^+\mathcal{S}_k[f^+]+\gamma^-\mathcal{S}_k[f^-]}{2}\begin{pmatrix}
1&0\\
0&1
\end{pmatrix}+\gamma^x \mathcal{S}_k[f^x]\begin{pmatrix}
1&0\\
0&0
\end{pmatrix}\\
&+\gamma^p \mathcal{S}_k[f^p]\begin{pmatrix}
0&0\\
0&1
\end{pmatrix}\, . 
\end{split}
\label{eq-sym-a}
\end{equation}
The Fourier transform of the anti-symmetric component, $B=(C-C^T)/2$ is instead
\begin{equation}
\begin{split}
&\hat{b}_{q_k}=\frac{\gamma_{q_k}}{2}\sigma\, , \\
&\gamma_{q_k}=\gamma^- \mathcal{S}_k[f^-]-\gamma^+\mathcal{S}_k[f^+]\, .
\end{split}
\label{eq-sym-b}
\end{equation}

\subsection{The covariance matrix}

For gaussian states, the system information is contained in the covariance matrix $G_{ij}=\langle \left\{r_i,r_j\right\}\rangle/2$ \cite{weedbrook2012,adesso2007,adesso2014}, where $\langle \cdot \rangle=\Tr(\rho \,\cdot)$ is the expectation on the quantum state $\rho$. 
Under the dynamics in Eq.~\eqref{Lindblad}, $G$ evolves, defining $X(t)=e^{t\, \Omega (2H+B)}$,  as \cite{heinosaari2010} (see also Appendix \ref{APP_Cov})
\begin{equation}
G(t)=X(t)G X^T(t)+\int_0^t du \, X(u)\Omega A\Omega^TX^T(u)\, .
\label{dyn-CV}
\end{equation}
From this equation, we can obtain the time-evolved covariance matrix in the space of quasi-momenta, just by applying the Fourier transform implemented by $U$. Using that $U$ is unitary, from Eq.~\eqref{dyn-CV} we obtain 
\begin{equation}
\hat{G}(t)=\hat{X}(t)\hat{G}(0)\hat{X}^\dagger(t)+\int_0^t du \, \hat{X}(u)\Omega \hat{A}\Omega^T \hat{X}^\dagger(u)
\end{equation}
where we have $\hat{X}(t)=e^{t\, \Omega(2\hat{H}+\hat{B})}$. Since the Fourier matrices $\hat{H},\hat{A},\hat{B}$ are all block-diagonal and assuming an initial translation-invariant state, we have that $\hat{G}(t)$ is block-diagonal with 
$$
\lceil\hat{G}(t)\rfloor_{kk}= \hat{g}_{q_k}(t)\, .
$$
The matrices $\hat{g}_{q_k}(t)$ evolves according to 
\begin{equation}
\hat{g}_{q_k}(t)=\hat{x}_{q_k}(t)\hat{g}_{q_k}(0)\hat{x}^\dagger_{q_k} (t)+\int_0^t du \, \hat{x}_{q_k}(u)\sigma  \hat{a}_{q_k}\sigma^T \hat{x}_{q_k}^\dagger(u)
\label{symbol-cv}
\end{equation}
where $\hat{x}_{q_k}(t)=e^{t\sigma(2\hat{h}_{q_k}+\hat{b}_{q_k})}$ is the symbol of the matrix $X(t)$.

In the following, we focus on the dynamics of  quantum entropies for a subsystem of $\ell$ adjacent sites embedded in such an open quantum many-body system. Our formula in Eq.~\eqref{formula} holds in the scaling limit of large $\ell$ and large times $t$, with $t/\ell$ fixed. In this limit, to observe a competition between coherent and irreversible effects, dissipation rates must be of order $\ell^{-1}$. For rates of order one, the system would  immediately converge to its steady-state, due to the large time limit. For rates weaker than $\ell^{-1}$, dissipation would be irrelevant. 

\section{Entropy from mixedness of the state} 
We start by deriving the entropy associated with the state of the quasi-particles being mixed, $s^{(n),\, {\rm mix}}_q$. To this end, we shall consider entropies of the full many-body state.

The R\'enyi-$n$ entropy is defined as $S^{(n)}=(1-n)^{-1}\ln \Tr \rho^n$ and the von Neumann entropy is included as the limiting case $n\to1$, yielding $S^{{\rm vN}}=-\Tr \rho \log \rho$. For gaussian states, these are computed exploiting the covariance matrix $G$  \cite{peschel2009}. To obtain the entropies one considers the matrix $\Sigma ={\rm i}\Omega G$. Its eigenvalues come into pairs $\pm \lambda_i$, with $\lambda_i\ge0$. Defining the function $y_\pm(x)=x\pm 1/2$, one then has
\begin{equation}
\begin{split}
&S^{(n)}=-\frac{1}{1-n}\sum_{i=1}^L \ln \left[y_+^n(\lambda_i)-y_-^n(\lambda_i)\right]\, , \\
&S^{{\rm vN}}=\sum_{i=1}^L\left[y_+(\lambda_i)\ln y_+(\lambda_i)-y_-(\lambda_i)\ln y_-(\lambda_i)\right]\, .
\end{split}
\label{Renyi-entropies}
\end{equation}

To find the contributions $s_{q}^{(n),\, {\rm mix}}$, we move to Fourier space where the matrix $\hat{G}(t)$ is block-diagonal, with blocks $\hat{g}_{q_k}(t)$ related to the  covariance matrix of the quasi-momentum $q_k$. Since $[U,\Omega]=0$, the eigenvalues of $\Sigma$ coincide with those of $\hat{\Sigma}=U\Sigma U^\dagger={\rm i}\Omega \hat{G}$. Thus, the full system entropy is given by the sum of all the contributions due to the different quasi-momenta, obtained from the positive eigenvalue  $\lambda_{q_k}(t)$ of ${\rm i}\sigma \hat{g}_{q_k}(t)$ as
\begin{equation}
s_{q_k}^{(n), \, {\rm mix}}(t)= -\frac{1}{1-n}\ln \left[y_+^n(\lambda_{q_k}(t)) -y_-^n(\lambda_{q_k}(t))\right]\, ,
\label{s_class_q}
\end{equation}
with  $y_\pm(x)=x\pm 1/2$. The entropy of the full system state is then given by
\begin{equation}
S^{(n)}(t)=\sum_{k=1}^L s_{q_k}^{(n), \, {\rm mix}}(t)\approx \frac{L}{2\pi}\int_0^{2\pi}\!\!dq\,s_{q}^{(n), \, {\rm mix}}(t)\, .
\label{s-class}
\end{equation}
The approximate behavior holds for $L\gg1$, in the continuum limit  for the quasi-momenta. From now on, when using $q$ instead of $q_k$ we will refer to quantities expressed in the continuous limit. 

\section{Entropy from quasi-particle densities} 
We now discuss the term $s^{(n),\, {\rm YY}}_{q}$, which solely depends on the quasi-particle density $\varrho_{q}$. For bosonic systems, this is given by  \cite{calabrese2018}
\begin{equation}
s^{(n),\, {\rm YY}}_{q}(t)=-\frac{1}{1-n}\ln \left[(\varrho_{q} (t)+1)^n-\varrho_{q}^n(t)\right]\, .
\label{s_quantum_q}
\end{equation}
For closed systems, quasi-particles are defined through the Hamiltonian eigenmodes and, thus, densities are time-independent. However, in our setting, the dynamical generator is not just given by the Hamiltonian but rather by the Lindblad map $\mathcal{L}$ in Eq.~\eqref{Lindblad}. It is thus natural to define quasi-particles as the ``eigenmodes" $\beta_{q},\beta_{q}^\dagger$ of $\mathcal{L}$. The densities are then obtained as $\varrho_{q}=\braket{n_q}$, where $n_{q}=\beta_{q}^\dagger \beta_q$ is the quasi-particle number operator. In general, these densities are time-dependent [cf.~Fig.~\ref{Fig1}(b)]. 

\begin{figure*}[t]
\centering
\includegraphics[width=\linewidth]{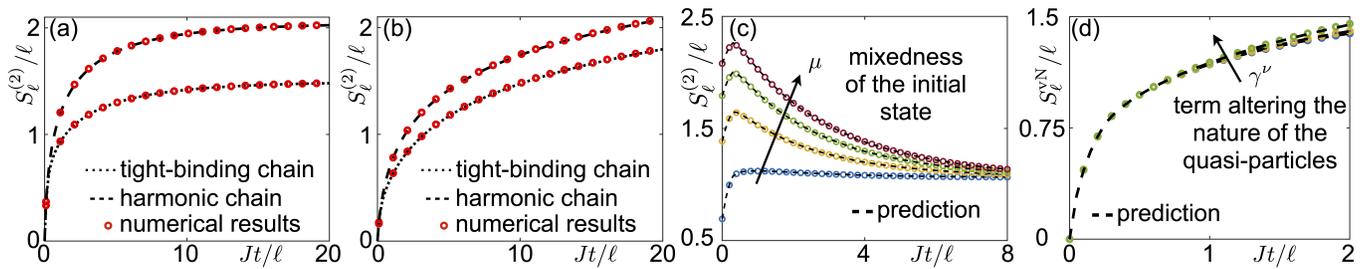}
\caption{{\bf R\'enyi and von Neumann entropy.} (a-b) Prediction of the R\'enyi entropy, for both the tight-binding model (dotted line) and the harmonic chain (dashed line), obtained from Eq.~\eqref{formula}. Red circles are exact numerical results. (a) Dissipative rates are $\gamma^{+}=\ell^{-1}/4$, $\gamma^{-}=\gamma^{p}=\gamma^x=\ell^{-1}$, while the correlation lengths of the processes (see main text) are $\xi^+=1$, $\xi^-=2$, $\xi^x=3$ and $\xi^p=4$. (b) We consider the ``critical" regime for which there is no steady state for the bosonic system. We achieve this by taking $\gamma^\pm=\ell^{-1}$ and $\gamma^{x/p}=0$. We have further set $\xi^{\pm}=1$. Plots in (a-b) are for $\ell=10$ and $L=100$. The initial state for the tight-binding model is the one in Eq.~\eqref{squeezed-thermal} for $\mu=\chi=1$, while for the harmonic chain it is the ground state of the model for $m_0=1$. (c) Tight-binding chain with non-local dissipation characterized by  $\gamma^-=4\gamma^+=\ell^{-1}$, $\gamma^x=\ell^{-1}/2$ and $\gamma^p=0.7\ell^{-1}$. Furthemore, we take $\xi^{\pm}=1$ and  $\xi^{x/p}=2$. The plot shows a comparison between numerical results for the Renyi-$2$ entropy of the subsystem  (circles) and our prediction in Eq.~\eqref{formula} for different initial mixed states [see Eq.~\eqref{squeezed-thermal}] parametrized by $\chi=1$ and $\mu=2,4,6,8$. We take $\ell=20$ and $L=300$. (d) Tight-binding chain with local decay and local pump of bosonic excitations with rates $\gamma^-=1.3\ell^{-1}$ and $\gamma^+=0.75\ell^{-1}$. Furthemore, we take the nearest-neighbor dissipative term proportional to $\gamma^\nu$, introduced in Eq.~\eqref{alt-QP}. We show the von Neumann entropy of the reduced state for $\gamma^\nu=0,0.2,0.4$. We compare our prediction (dashed line) with exact numerical results (circles). We take $\ell=40$ and $L=400$ and an initial state with $\mu=\chi=1$.  }
\label{Fig2}
\end{figure*}  

\subsection{Eigenmodes of the Lindblad generator}
To find the Lindblad eigenmodes, one should find annihilation and creation operators, $[\beta_{q},\beta_{p}^\dagger]=\delta_{qp}$, such that 
$$
\mathcal{L}[\beta_q]=\zeta_q \beta_q\, ,\qquad \mbox{ which implies } \, \qquad \mathcal{L}[\beta_q^\dagger ]=\zeta_q^* \beta^\dagger_q\, .
$$

The starting point is to compute the action of the Lindblad operator on quadrature operators, 
\begin{equation}
\mathcal{L}[r_i]= \sum_{j}[\Omega (2H +B)]_{ij}r_j\, .
\label{eq-cov-s2}
\end{equation}
Then, we transform the operators $r$ in Fourier space by defining the vector $\hat{r}=Ur$. This needs to be understood as the vector of elements $\hat{r}_{i}=\sum_{j=1}^{2L}U_{ij}r_j$. Using the form of $U$, we can actually define 
\begin{equation}
\begin{split}
&\hat{r}_{2k-1}=\varphi_{q_k}:=\frac{1}{\sqrt{L}}\sum_{j=1}^L e^{{\rm i} q_k j} x_{j}\, , \\
&\hat{r}_{2k}=\pi_{q_k}:=\frac{1}{\sqrt{L}}\sum_{j=1}^L e^{{\rm i} q_k j} p_{j}\, ,
\end{split}
\label{eq-Fourier-quad}
\end{equation}
for $k=1,2,3,\dots L$. Recall that $q_k=2\pi k /L$ is the quasi-momentum. Exploiting Eq. \eqref{eq-cov-s2}, we obtain 
\begin{equation}
\begin{split}
\mathcal{L}[\hat{r}_i]&=\sum_{j=1}^{2L}\left\{U\left[\Omega\left(2H+B\right)\right]U^\dagger \right\}_{ij} \hat{r}_j\\
&= \sum_{j=1}^{2L}\left[\Omega\left(2\hat{H}+\hat{B}\right)\right]_{ij} \hat{r}_j\, .
\end{split}
\label{eq-eig-s1}
\end{equation}
For the last equality, we have used that $[U,\Omega]=0$ as well as $\hat{H}=UHU^\dagger $ and $\hat{B}=UBU^\dagger$. Since $\hat{H}$ and $\hat{B}$ are block diagonal, we can ``unravel" Eq.~\eqref{eq-eig-s1} into $L$ relations involving $2\times2$ matrices. These read as 
\begin{equation}
\mathcal{L}\left[\begin{pmatrix}
\varphi_{q_k}\\ 
\pi_{q_k}
\end{pmatrix}\right]=\sigma\left(2\hat{h}_{q_k}+\hat{b}_{q_k}\right)\begin{pmatrix}
\varphi_{q_k}\\ 
\pi_{q_k}
\end{pmatrix}\, , 
\label{eq-eig-s2}
\end{equation}
and with $\hat{h}_{q_k},\hat{b}_{q_k}$ being the symbols of $H,B$. To find the eigenmodes, we need to find a linear combination $\beta_{q_k}$ of the operators $\varphi_{q_k},\pi_{q_k}$, such that $\mathcal{L}[\beta_{q_k}]\propto \beta_{q_k}$. Since Eq.~\eqref{eq-eig-s2} contains the term $\hat{b}_{q_k}$, in general, the presence of dissipation is expected to modify the structure of the eigenmodes due to the Hamiltonian contribution only.

Together with the eigenmodes, one also obtains the eigenvalues $\zeta_q$ associated with them. In full generality one has 
\begin{equation}
\mathcal{L}\left[\beta_{q}\right]=-\left(\frac{\gamma_{q}}{2} +{\rm i} \, {\rm e}_{q} \right)\beta_{q}\, ,
\label{eig_mod}
\end{equation}
with $\gamma_q$ and $\rm{e}_q$ real. Here, $\gamma_{q}$ is the ``decay" rate (it can be negative for bosons) for the $q$th quasi-particles. The function ${\rm e}_{q}$ plays the role of a dispersion relation and, in analogy with closed systems, it provides the quasi-particle velocity as $v_{q}={\rm e}'_q$. Interestingly, we note that for the dissipative processes mentioned above, the eigenmodes in Eq.~\eqref{eig_mod} coincide with those of the Hamiltonian. This is due to the fact that, for gain/loss and diffusion dissipation, the matrix $\sigma\hat{b}_{q_k}$ is proportional to the identity [see definition of $\hat{b}_{q_k}$ in Eq.~\eqref{eq-sym-b}], so that the  eigenmodes of the Hamiltonian are also eigenmodes of the Lindblad generator. However, our approach is by no means limited to these cases and also applies to instances in which  dissipation alters the nature of the Hamiltonian quasi-particles (see an example below).

As a consequence of Eq.~\eqref{eig_mod}, and also of the relation in Eq.~\eqref{eq-cov-s1},  the operator $n_q$ obeys
$\dot{n}_q=\mathcal{L}\left[n_{q}\right]=-\gamma_{q} n_{q}+\alpha_{q}$,
with $\alpha_{q}\ge0$ a positive constant which solely depends on the structure of the dissipation. By integrating this equation, we find 
\begin{equation}
\varrho_{q}(t) =e^{-t\, \gamma_{q} }\varrho_{q}(0)+\frac{\alpha_{q}}{\gamma_{q}}\left(1-e^{-t\, \gamma_{q}}\right)\, ,
\label{density}
\end{equation}
where $\varrho_{q}(0)$ are the densities in the initial quantum state. Their  dynamics affects $s^{(n),\, {\rm YY}}_{q}$ through Eq.~\eqref{s_quantum_q}.

\section{Applications} 
So far, we have presented the different terms appearing in Eq.~\eqref{formula} and we have shown how to derive them for quadratic open quantum systems. We can thus now benchmark our formula against numerical simulations. To this end, we will consider two different Hamiltonian models subject to the dissipative processes discussed above in different combinations.

\subsection{Tight-binding chain}
As a first example, we look at a tight-binding bosonic hopping model defined by the  Hamiltonian
\begin{equation}
E=J\sum_{i=1}^L\left(a_ia_{i+1}^\dagger+a_i^\dagger a_{i+1}\right)\, ,
\label{bos-hop}
\end{equation}
where $a_i=(x_i+{\rm i}p_i )/\sqrt{2}$ is the annihilation operator for site $i$. Expanding this in the quadrature operators, this Hamiltonian gives rise to a matrix $H$, with form given in Eqs.~\eqref{circ_M}-\eqref{M} and just the matrix $h_1$ different from zero and equal to 
$$
h_1=\frac{J}{2}\begin{pmatrix}
1&0\\
0&1
\end{pmatrix}\, .
$$
The eigenmodes $\beta_{q_k}$ of the Hamiltonian are given by 
$$
\beta_{q_k}=\frac{1}{\sqrt{L}}\sum_{i=1}^L e^{{\rm i}q_k}a_i\, ,
$$
with dispersion relation and quasi-particle velocities 
$$
{\rm e}_{q_k}=2\cos (q_k)\, , \quad \mbox{ and } \quad v_{q_k}=-2\sin (q_k)\, .
$$

We consider  the dissipative processes introduced above, with $f^\alpha(d)=e^{-d/\xi_{\alpha}}$, where $\xi_\alpha$ encode how the  non-local dissipative processes are ``correlated" in space. The action of the Lindblad on the density of quasi-particles, $n_{q_k}=\beta_{q_k}^\dagger \beta_{q_k}$, can be computed, using Eqs.~\eqref{eig_mod}-\eqref{eq-cov-s1}, and is given by 
$$
\mathcal{L}[n_{q_k}]=-\gamma_{q_k} n_{q_k}+\alpha_{q_k}\, ,
$$
with $\gamma_{q_k}$ given in Eq.~\eqref{eq-sym-b} and 
$$
\alpha_{q_k}=\gamma^+\mathcal{S}_k[f^+]+\frac{\gamma^x}{2}\mathcal{S}_k[f^x]+\frac{\gamma^p}{2}\mathcal{S}_k[f^p]\, .
$$
As initial state we take the one described by the block-diagonal covariance matrix 
\begin{equation}
\lceil G_\mu\rfloor_{ii}= \frac{\mu}{2}\begin{pmatrix}
e^{\chi}&0\\
0&e^{-\chi}
\end{pmatrix}\, ,
\label{squeezed-thermal}
\end{equation}
where $\mu\ge 1$. This covariance matrix is associated with a squeezed thermal state and allows us to show how our formula are also valid for initial mixed states. In the above expression, $\chi$ is the squeezing parameter while the parameter $\mu$ represents the average density in the thermal state $\rho\propto e^{-1/T \sum_{i=1}^L a^\dagger_i a_i}$.

\subsection{Harmonic chain}
As a second example, we consider the harmonic chain
\begin{equation}
E=\frac{J}{2}\sum_{i=1}^L\left(p_i^2+m^2 x_i^2+(x_i-x_{i+1})^2\right)\, .
\end{equation}
We take as initial state the ground state of the Hamiltonian for $m=m_0$ \cite{coser2014,calabrese2018}. The system dynamics is characterized by non-local dissipative terms, as for the previous model, and by a quenched value of $m\neq m_0$.  

Such a Hamiltonian gives rise to a matrix $H$ of the form in Eqs.~\eqref{circ_M}-\eqref{M} with 
$$
h_0=J\begin{pmatrix}
\frac{m^2}{2}+1&0\\
0&\frac{1}{2}
\end{pmatrix}\, ,\qquad \mbox{ and } \qquad h_1=-\frac{J}{2}\begin{pmatrix}
1&0\\
0&0
\end{pmatrix}\, .
$$
The eigenmodes of $H$ are given in terms of the Fourier operators \cite{coser2014,calabrese2018}
\begin{equation*}
\begin{split}
&\beta_{q_k}=\frac{1}{\sqrt{2 {\rm e}_{q_k}}}\left({\rm e}_{q_k}\varphi_{q_k}+{\rm i}\pi_{q_k}\right)\, , \\ 
&\beta_{q_k}^\dagger =\frac{1}{\sqrt{2 {\rm e}_{q_k}}}\left({\rm e}_{q_k}\varphi_{-q_k}-{\rm i}\pi_{-q_k}\right)\, ,
\end{split}
\end{equation*}
where $\varphi_{-q_k}=\varphi_{q_k}^\dagger$, $\pi_{-q_k}=\pi_{q_k}^\dagger$. The dispersion relation ${\rm e}_{q_k}$ and the quasi-particle velocities are 
\begin{equation*}
\begin{split}
&{\rm e}_{q_k}=J\sqrt{m^2+2[1-\cos(q_k)]}\, , \\ 
&v_{q_k}=\frac{J\sin (q_k)}{\sqrt{m^2+2[1-\cos(q_k)]}}\, .
\end{split}
\end{equation*}

The action of the full Lindblad generator on the eigenmode $\beta_{q_k}$ gives 
$$
\mathcal{L}[\beta_{q_k}]=-\left(\frac{\gamma_{q_k}}{2}+{\rm i}{\rm  e}_{q_k}\right)\beta_{q_k}\, ,
$$
where $\gamma_{q_k}$ is given in Eq.~\eqref{eq-sym-b}. On the number operator  $n_{q_k}=\beta_{q_k}^\dagger \beta_{q_k}$ it gives 
$$
\mathcal{L}[n_{q_k}]=-\gamma_{q_k}n_{q_k}+\alpha_{q_k}\, ,
$$
with 
\begin{equation}
\begin{split}
\alpha_{q_k}=&\gamma^- \mathcal{S}_k[f^-]\frac{(1-{\rm e}_{q_k})^2}{4 {\rm e}_{q_k}}+\gamma^+ \mathcal{S}_k[f^+]\frac{(1+{\rm e}_{q_k})^2}{4 {\rm e}_{q_k}}\\
&+\frac{\gamma^x}{2{\rm e}_{q_k}} \mathcal{S}_k[f^x]+\frac{\gamma^p {\rm e}_{q_k}}{2} \mathcal{S}_k[f^p]\, .
\end{split}
\end{equation}

\subsection{Numerical checks on the entropy}

In Fig.~\ref{Fig2}(a-b), we show a comparison between the prediction obtained through Eq.~\eqref{formula} for the dynamics of the R\'enyi-$2$ entropy and exact numerical results, for both models. The agreement is remarkable also in regimes in which $\gamma_{q}=0$ and the entropy increases logarithmically with time, $S_\ell^{(2)}\approx \ell \ln t$ [shown in Fig.~\ref{Fig2}(b)]. Moreover, in Fig.~\ref{Fig2}(c), we compare our predicition in Eq.~\eqref{formula} with numerical results for the tight-binding model when starting from the mixed state obtained from Eq.~\eqref{squeezed-thermal} with $\mu>1$. Also in this case the prediction is satisfactory and numerical results tend to it in the scaling limit employed. 

\subsection{Mutual information}

The entropies $S_\ell^{(n)}$ do not quantify correlations, due to the presence of contributions from the mixedness (statistical uncertainty) of the full many-body state [cf.~Fig.~\ref{Fig1}(b)]. However, building on our interpretation of Eq.~\eqref{formula} in terms of a dissipative quasi-particle picture, we can still achieve a description of correlations between the subsystem of interest and the remainder. Recalling Eq.~\eqref{s-class}, we identify the contribution due to statistical uncertainty with the first term in Eq.~\eqref{formula}. Thus, subtracting the latter term to the subsystem entropy $S_\ell^{(n)}$, we define the quantity
\begin{equation}
S_{\ell, {\rm pairs}}^{(n)}=\int \frac{dq}{2\pi}\min(2|v_q|t,\ell)\!\left[s_q^{(n), \, {\rm YY}}(t)-s_q^{(n),\,{\rm mix}}(t)\right]\, .
\label{propag}
\end{equation}
Such a contribution is only sensitive to quasi-particle pairs which are shared by the subsystem and the remainder. As such, it must be invariant under exchange of  these two parts, i.e.~$S_{\ell, {\rm pairs}}^{(n)}= S_{L-\ell, {\rm pairs}}^{(n)}$, at leading order in $\ell$ and in our scaling limit \cite{maity2020,alba2021}. This relation suggests that, in this limit, we can compute the mutual information, given by $\mathcal{I}^{(n)}=S^{(n)}_\ell+S^{(n)}_{L-\ell}-S^{(n)}$, as $\mathcal{I}^{(n)}=2S_{\ell, {\rm pairs}}^{(n)}$. In Fig.~\ref{Fig3}(a), we show our prediction $S_{\ell, {\rm pairs}}^{(2)}$ for the tight-binding chain starting from the mixed state in Eq.~\eqref{squeezed-thermal}. In Fig.~\ref{Fig3}(b), we show instead the prediciton for $S_{\ell, {\rm pairs}}^{\rm vN}$ for the harmonic chain. We compare both predicitons with exact numerical results for the mutual information between the subsystem and the remainder of the many-body system. As shown in the insets, the extrapolation of the numerical results  (bullets) converges to our prediction (square), showing how indeed the formula in Eq.~\eqref{propag} provides the behavior of quantum correlations in the system.

\section{Quasi-particles altered by dissipation}  
In the previous examples, we have considered dissipative processes which, as discussed after Eq.~\eqref{eig_mod}, preserve the nature of the Hamiltonian quasi-particles. To demonstrate the generality of our approach, we now show that our formulae Eq.~\eqref{formula}-\eqref{propag}  remain valid beyond these instances. 

To this end, we consider again the tight-binding model in Eq.~\eqref{bos-hop}, subject to local pump and decay of excitations, i.e.~with functions $f^{\pm}_{ij}=\delta_{ij}$. In addition, we introduce a dissipative contribution described by the matrix \begin{equation}
\lceil B^\nu\rfloor_{ij}=\gamma^\nu f^\nu_{ij} \, b^\nu\, , \quad b^\nu=\begin{pmatrix}
1&0\\
0&0
\end{pmatrix}\, ,
\label{alt-QP}
\end{equation}
with $f^\nu$ being the anti-symmetric function $f_{ij}^\nu=\delta_{1,{\rm mod}(j-i,L)}-\delta_{L-1,{\rm mod}(j-i,L)}$, solely involving nearest-neighboring dissipative ``coupling". Namely, the matrix $B^\nu$ is of the form given in Eqs.~\eqref{circ_M} with $b^\nu_1=\gamma^\nu b^\nu$ and $b^\nu_{L-1}=-\gamma^\nu b^\nu$. The matrix $C$ in this case is thus $C=C^++C^-+{\rm i}B^\nu$, and all rates must be chosen such that $C\ge0$.

While the term $s_{q}^{(n), \, {\rm mix}}$ is straightforwardly given by Eq.~\eqref{s_class_q}, in order to exploit our formulae the main challenge is to find the quasi-particles of such a dissipative dynamics. 

\begin{figure*}[t]
\centering
\includegraphics[width=\linewidth]{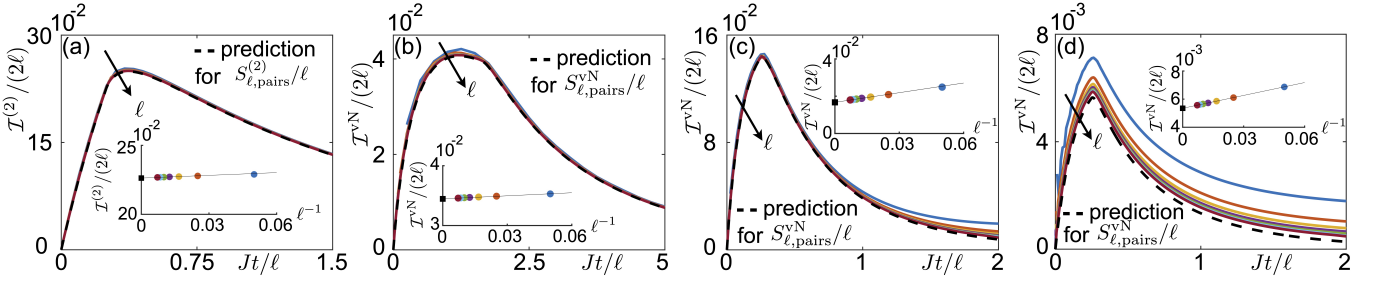}
\caption{{\bf Mutual information.} Comparison between numerical results for (half of) the mutual information $\mathcal{I}^{(n)}$ and the prediction for $S_{\ell, {\rm pairs}}^{(n)}$ in Eq.~\eqref{propag}. (a)  Renyi-$2$ mutual information for the tight-binding chain. We compare numerical results for (half of) the mutual information and our prediction for $S^{(2)}_{\ell,{\rm pairs}}$ in Eq.~\eqref{propag}. The initial state is the one in Eq.~\eqref{squeezed-thermal} with $\mu=5$, other parameters are as in Fig.~\ref{Fig2}(c). The inset shows how the extrapolation of the numerical results (bullets) to $\ell\to\infty$ matches our prediction (square) for $Jt/\ell=0.25$. (b) Harmonic chain: we consider non-local decay with $\gamma^-=\ell^{-1}/2$, $\xi^-=2$ and $m=3$. The initial state is the ground state for $m_0=1$. Inset: the extrapolation of the numerical results (bullets) to $\ell\to\infty$ matches the prediction (square) for $Jt/\ell=2$. (c) Tight-binding chain subject to local decay ($\gamma^-=1.3\ell^{-1}$) and pump ($\gamma^+=0.75\ell^{-1}$) of excitations and to the term introduced in Eq.~\eqref{alt-QP} ($\gamma^\nu=0.4\ell^{-1}$) which modifies the nature of the Hamiltonian quasi-particles. The initial state is the squeezed vacuum with $\chi=1$.  The inset shows the agreement between numerical results (bullets) and our prediction (square) for $Jt/\ell=1.5$. (d) Ising model: we consider decay (see details in Appendix \ref{App_Ising}) with $\gamma^-=\ell^{-1}/2$, $\xi^-=2$ and Hamiltonian parameter $h^x=5$. The initial state is the ground state for $h_0^x=3$. The inset shows the agreement between extrapolation of numerical results for (half of) the von Neumann mutual information (bullets) and our prediction (square) for $Jt/\ell=0.3$. For all panels, we have considered $\ell=20,40,60,\dots 140$ and $L$ large enough to avoid finite-size effects.}
\label{Fig3}
\end{figure*} 

In order to find the eigenmodes of the Lindblad generator identified by the matrix $C$ of this section, we consider the analogue of Eq.~\eqref{eq-eig-s2} (moving to Fourier space), which for this case becomes
\begin{equation}
\mathcal{L}\left[\begin{pmatrix}
\varphi_{q_k}\\ 
\pi_{q_k}
\end{pmatrix}\right]=\sigma\left(2\hat{h}_{q_k}+\hat{b}_{q_k}+\hat{b}^\nu_{q_k}\right)\begin{pmatrix}
\varphi_{q_k}\\ 
\pi_{q_k}
\end{pmatrix}\, , 
\label{eq-eig-alt-s1}
\end{equation}
with
\begin{equation}
\begin{split}
&\hat{h}_{q_k}=J\begin{pmatrix}
\cos q_k&0 \\ 
0&\cos q_k
\end{pmatrix}\, , \qquad \hat{b}_{q_k}=\begin{pmatrix}
0&\gamma/2 \\ 
-\gamma/2&0
\end{pmatrix}\, , \\
&\hat{b}^\nu_{q_k}=\begin{pmatrix}
-2{\rm i}\gamma^\nu\sin q_k&0 \\ 
0&0
\end{pmatrix}\, .
\end{split}
\label{eq-eig-alt-s2}
\end{equation}
In the above equations, we have $\gamma=\gamma^--\gamma^+$, which does not depend on $q_k$ since decay and pump processes are local (i.e.~diagonal). Computing the product between the matrices in Eq.~\eqref{eq-eig-alt-s1} we find, 
\begin{equation}
\mathcal{L}\left[\!\begin{pmatrix}
\varphi_{q_k}\\ 
\pi_{q_k}
\end{pmatrix}\!\right]\!=\!\begin{pmatrix}
-\gamma/2&2J\cos q_k\\
-2J\cos q_k+2{\rm i}\gamma^\nu\sin q_k&-\gamma/2\end{pmatrix}\!\!
\begin{pmatrix}
\varphi_{q_k}\\ 
\pi_{q_k}
\end{pmatrix}.
\label{eq-eig-alt-s3}
\end{equation}
The fact that the contribution proportional to $\gamma^\nu$ modifies the quasi-particles of the Hamiltonian manifests in the fact that the matrix $\sigma \hat{b}^\nu_{q_k}$ does not commute with the matrix $\sigma \hat{h}_{q_k}$. We now note that while $J$ is of order one, $\gamma^\nu$, just like all other dissipation rates in our scaling limit, is of order $\ell^{-1}$. Therefore, in order to find how the Hamiltonian quasi-particles are altered by dissipation, we can proceed by finding the eigenmodes of $\mathcal{L}$ through a first-order perturbation theory in $\gamma^\nu$. (Also $\gamma$ is proportional to $\ell^{-1}$ and thus small in the scaling limit considered. However, we do not have to expand in $\gamma$ since decay and pump processes do not modify the Hamiltonian eigenmodes.)

We thus proceed as follows. We decompose the generator as $\mathcal{L}=\mathcal{L}_0+\gamma^\nu \mathcal{L}_1$, where the first term is determined by the tight-binding Hamiltonian plus pump and decay dissipation, while $\mathcal{L}_1$ is solely determined by the contribution proportional to the rate $\gamma^\nu$. Similarly, we  decompose the eigenmodes of $\mathcal{L}$ as $\beta_{q_k}=\beta_{q_k}^0+\gamma^\nu \beta^1_{q_k}$. The modes $\beta_{q_k}^0$ are the eigenmodes of $\mathcal{L}_0$, coinciding with those of the tight-binding Hamiltonian, 
$$
\mathcal{L}_0[\beta_{q_k}^0]=\zeta_0\beta_{q_k}^0=\left(-\frac{\gamma}{2}- {\rm i e}_{q_k}\right)\beta_{q_k}^0\, .
$$
The term $\beta_{q_k}^1$ represents the correction due to the dissipative contribution modifying  the quasi-particles, which we can generically write as
$$
\beta_{q_k}^1=u_1\varphi_{q_k}+{\rm i} u_2\pi_{q_k}\,;
$$
$u_1,u_2$ are two complex parameters that need to be determined. A first constraint comes from asking that $[\beta_{q_k},\beta_{q_k}^\dagger]=1$ up to first-order in $\gamma^\nu$. This gives
\begin{equation}
\begin{split}
[\beta_{q_k},\beta_{q_k}^\dagger]&=[\beta_{q_k}^0,(\beta_{q_k}^0)^\dagger]\\&+\gamma^\nu \left(\left[\beta_{q_k}^1,(\beta_{q_k}^0)^\dagger\right]
+\left[\beta_{q_k}^0,(\beta_{q_k}^1)^\dagger\right]\right)+\dots =1\, ,\end{split}
\end{equation}
which can be shown to be satisfied if ${\rm Re}(u_1+u_2)=0$. We thus set $u_1=s+iv_1$ and $u_2=-s+iv_2$, with $s,v_1,v_2$ real parameters.

We now derive the perturbative equations which will provide constraints for $s$, $v_1$ and $v_2$. Expanding the eigenvalue relation $\mathcal{L}[\beta_{q_k}]=\zeta \beta_{q_k}$ up to first order in $\gamma^\nu$ and using $\zeta=\zeta_0+\gamma^\nu \zeta_1$, we find 
$$
\mathcal{L}_0[\beta_{q_k}^0]+\gamma^\nu \mathcal{L}_1[\beta_{q_k}^0]+\gamma^\nu \mathcal{L}_0[\beta_{q_k}^1]\!=\!\zeta_0 \beta_{q_k}^0\!+\!\gamma^\nu \zeta_0 \beta_{q_k}^1\!+\!\gamma^\nu \zeta_1\beta_{q_k}^0\, .
$$
Calculating the action of the different parts of the generator [using Eq.~\eqref{eq-eig-alt-s3}] on the various operators, and simplifying several terms, we find that the following equality must be satisfied 
\begin{equation}
\begin{split}
&\varphi_{q_k}\Bigg\{\left[2(v_2-v_1)J\cos q_k-\sqrt{2}\sin q_k-\frac{\zeta_1^{\rm Re}}{\sqrt{2}}\right]\\
&+i\left[4 sJ\cos q_k -\frac{\zeta_1^{\rm Im}}{\sqrt{2}}\right]\Bigg\}+\\
&+\pi_{q_k}\Bigg\{\left[4 sJ\cos q_k +\frac{\zeta_1^{\rm Im}}{\sqrt{2}}\right]\\
&+i\left[2(v_1-v_2)J\cos q_k-\frac{\zeta_1^{\rm Re}}{\sqrt{2}}\right]\Bigg\}=0\, ,
\end{split}
\end{equation}
with $\zeta_1^{\rm Re}$ and $\zeta_1^{\rm Im}$ being the real and the imaginary part of the term $\zeta_1$. Now, the aim is to find a combination of $s,v_1,v_2,\zeta_1$ for which all square brackets in the above equation vanish. The solution can be found and gives $s=0$, $(v_2-v_1)=\sqrt{2}\sin q_k/(4J\cos q_k)$, as well as $\zeta_1^{\rm Im}=0$ and $\zeta_1^{\rm Re}=-\sin q_k$. 

The above correction to the eigenvalue is of extreme importance since it gives the proper decay rate for the altered quasi-particles [cf.~Eq.~\eqref{eig_mod}]
\begin{equation}
\mathcal{L}[\beta_{q_k}]=\left(-\frac{\gamma}{2}-\gamma^\nu\sin q_k-{\rm i e}_{q_k}\right)\beta_{q_k}\, .
\label{eq-eig-alt}
\end{equation}
We now can find the rate equation applying the Lindblad generator on the quasi-particle number operator $\beta_{q_k}^\dagger \beta_{q_k}$. As done also in Eq.~\eqref{eq-cov-s1}, we can write this as
$$
\mathcal{L}[\beta_{q_k}^\dagger \beta_{q_k}]=\mathcal{L}[\beta_{q_k}^\dagger ]\beta_{q_k}+\beta_{q_k}^\dagger\mathcal{L}[ \beta_{q_k}]+\!\sum_{i,j=1}^{2L}\!\!C_{ij}[r_i,\beta_{q_k}^\dagger][\beta_{q_k},r_j].
$$
For the first two terms on the right-hand-side of the above equation, we can readily use the result in Eq.~\eqref{eq-eig-alt}. Noticing that $C$ is already of order $\ell^{-1}$ (since both $\gamma^\pm$ and $\gamma^\nu$ are of order $\ell^{-1}$) the last term in the above equation can be determined by neglecting the correction to the eigenmodes $\beta_{q_k}^1$. As such we have
\begin{equation}
\begin{split}
\sum_{i,j=1}^{2L}C_{ij}[r_i,\beta_{q_k}^\dagger][\beta_{q_k},r_j]&\approx \sum_{i,j=1}^{2L}C_{ij}[r_i,(\beta_{q_k}^0)^\dagger][\beta_{q_k}^0,r_j]\\
&=\gamma^+-\gamma^\nu \sin q_k\, ,
\end{split}
\end{equation}
and the rate equation reads
\begin{equation}
\dot{\varrho}_{q_k}(t)=-(\gamma +2\gamma^\nu\sin q_k)\varrho_{q_k}(t)+\gamma^+-\gamma^\nu \sin q_k\, .
\label{rate-altered}
\end{equation}
For the initial value of $\varrho_{q_k}(0)$, we can again use the fact that the correction term $\beta_{q_k}^1$ is of order $\ell^{-1}$ so that $\varrho_{q_k}(0)=\langle \beta_{q_k}^\dagger \beta_{q_k}\rangle \approx \langle (\beta_{q_k}^0)^\dagger \beta_{q_k}^0\rangle$, in the large $\ell$ limit, where the expectation is computed with the initial state. We recall here that the dissipative rates in Eq.~\eqref{rate-altered} remain relevant, even though of order $\ell^{-1}$, since time is rescaled by $\ell$ so that the product $\gamma^{\pm/\nu} t$ remains finite (see also discussion on the scaling limit at the end of Section \ref{sec_Lind}). We note that the above perturbation theory developed for this example is actually very general and can be used to find the eigenmodes for any quadratic Lindblad generator in our scaling limit.

In Fig.~\ref{Fig2}(d), we show that our formula in Eq.~\eqref{formula} correctly captures the subsystem entropy also for the model described here. Moreover, in Fig.~\ref{Fig3}(c), we also show how exact numerical results for the mutual information tend towards our prediction in Eq.~\eqref{propag}, as the system size is increased.

\section{Discussion} We introduced two key formulae [Eq.~\eqref{formula} and Eq.~\eqref{propag}] which describe the time-evolution of subsystem entropies and of correlations ---through the mutual information--- in generic quadratic open quantum systems. These formulae encode a dissipative quasi-particle picture which is predicated on the existence of pairs of propagating dissipative quasi-particle excitations.

For the sake of concreteness, we discussed in detail the most common dissipative processes [cf.~Eq.~\eqref{com-proc}]. For these, we found that the eigenmodes of the Lindblad dynamics ---defining the quasi-particles--- are solely determined by the Hamiltonian. However, our formulae remain valid also when the nature of the quasi-particles changes in the presence of dissipation. As we have shown, even in these instances, we can compute the entropy contributions in Eq.~\eqref{s_class_q}, find the eigenmodes of the Lindblad generator and obtain the rate equation for the densities $\varrho_{q}(t)$. This demonstrates the generality of our approach as well as its effectiveness in studying correlations in open quantum many-body systems.

In support of the broad applicability of our approach, we finally mention that this holds also for fermionic systems. We provide an example in Appendix \ref{App_Ising} [see Fig.~\ref{Fig3}(d)] and we further refer to the recent Ref.~\cite{alba2021c} for a derivation of Eq.~\eqref{formula} in the case of the quadratic dissipative Ising model.

\section*{Acknowledgements} 
F.C.~acknowledges support from the “Wissenschaftler-R\"uckkehrprogramm GSO/CZS” of the Carl-Zeiss-Stiftung and the German Scholars Organization e.V., as well as through the Deutsche Forschungsgemeinsschaft (DFG, German Research Foundation) under Project No. 435696605. V.A.~acknowledges support from the European Research Council under ERC Advanced grant No. 743032 DYNAMINT.

\appendix

\section{Time-Evolution of the covariance matrix}
\label{APP_Cov}
For completeness, we illustrate here the main steps to obtain the time-dependence of the covariance matrix in Eq. \eqref{dyn-CV}. This is completely established by the dynamics of all possible two-point operators $r_k r_h$.

We start with some considerations on the Lindblad operator in Eq. \eqref{Lindblad}. This is equivalent to 
$$
\mathcal{L}[X]=i[E,X]+\sum_{i,j=1}^{2L}\frac{C_{ij}}{2}\left([r_i,X]r_j +r_i[X,r_j]\right)\, ,
$$
and one can show that 
\begin{equation}
\mathcal{L}[XY]=X\mathcal{L}[Y]+\mathcal{L}[X]Y+\sum_{i,j=1}^{2L}C_{ij}[r_i,X][Y,r_j]\, .
\label{eq-cov-s1}
\end{equation}
This is useful to evaluate the action of the Lindblad operator on quadratic operators. To this end, we first compute the action of the Lindblad on linear operators,
\begin{equation}
\mathcal{L}[r_i]=\sum_{j=1}^{2L}\left[\Omega(2H+B)\right]_{ij}r_j\, .
\label{eq-cov-s2bis}
\end{equation}
Furthermore, the last term in Eq. \eqref{eq-cov-s1} is proportional to the identity and reads as 
$$
\sum_{i,j=1}^{2L}C_{ij}[r_i,r_k][r_h,r_j]=\left[\Omega C\Omega^T \right]_{kh}\, .
$$
We now define the $2L\times 2L$ matrix $\Gamma_{kh}=\langle r_k r_h\rangle$. Its  derivative is determined by the Lindblad generator as 
$$
\frac{d}{dt}\Gamma_{kh}(t)=\langle \mathcal{L}[r_kr_h]\rangle_t\, .
$$ 
Using the above results, we find that
$$
\frac{d}{dt}\Gamma(t)=[\Omega(2H+B)]\Gamma(t)+\Gamma(t)[\Omega(2H+B)]^T+\Omega C\Omega^T\, . 
$$
Noticing that the covariance matrix $G$ is nothing but $G=(\Gamma +\Gamma^T)/2$, we find
$$
\frac{d}{dt}G(t)=[\Omega(2H+B)]G(t)+G(t)[\Omega(2H+B)]^T+\Omega A\Omega^T\, , 
$$
whose solution is the one reported in Eq.~\eqref{dyn-CV}.

\section{Example for a fermionic system}
\label{App_Ising}

Our formulae Eq.~\eqref{formula} and Eq.~\eqref{propag} are also valid for fermionic quadratic open quantum systems. In this case, the vector $r$ contains Majorana operators $r=(w_1^1,w_1^2,w_2^1,w^2_2,\dots w_L^1,w_L^2)$, obeying the anticommutation relations $\{r_i,r_j\}=2\delta_{ij}$. For gaussian states, the full information is encoded in the (fermionic) covariance matrix $G_{ij}=\braket{[r_i,r_j]}/2$. The two entropic contributions $s_q^{(n),\, {\rm mix}},s_q^{(n),\, {\rm YY}}$ can be determined following the steps presented in the main text and using fermionic relations for the dynamics of the covariance matrix \cite{kos2017} and the analogue of Eqs.~\eqref{Renyi-entropies},\eqref{s_class_q},\eqref{s_quantum_q} for fermions \cite{alba2021}. The generator $\mathcal{L}$ can be written in a form similar to Eq.~\eqref{Lindblad} \cite{prosen2008}. For instance, the pump and loss dissipative matrices are in this case specified by the blocks
$$
c^\pm=\gamma^{\pm}\begin{pmatrix}
1&\pm {\rm i}\\ \mp {\rm i}&1
\end{pmatrix}\, .
$$.

In the following, we work out an exampe showing the validity of our approach. 

\subsection{(Fermionic) Transverse field Ising chain}
We consider the transverse field Ising model in its formulation with Majorana operators. 
The Hamiltonian is 
$$
E=-{\rm i}J\sum_{i=1}^Lw^2_i w^1_{i+1}+{\rm i}Jh^x\sum_{i=1}^Lw_i^2w_i^1={\rm i}\sum_{i,j=1}^{2L}H_{ij}r_ir_j\, .
$$
The matrix $H$ is an anti-symmetric matrix and has the form 
$$
H=\begin{pmatrix}
h_0&h_1&0&&\dots &-h_1^T\\
-h_1^T&h_0&h_1&0&&\vdots\\
&-h_1^T&h_0&h_1&\ddots\\
\vdots & \ddots &\ddots & \ddots& &0\\
&&&-h_1^T&h_0&h_1\\
h_1&\dots &&&-h_1^T&h_0
\end{pmatrix}\, ,
$$
with 
$$
h_0=\frac{Jh^x}{2}\begin{pmatrix}
0&-1\\
1&0
\end{pmatrix}\, , \qquad \mbox{ and }\qquad h_1=\frac{J}{2}\begin{pmatrix}
0&0\\
-1&0
\end{pmatrix}\, .
$$
By applying the Hamiltonian part of the generator on linear operators, we obtain
$$
i[E,r_i]=4\sum_{j=1}^{2L}H_{ij}r_j\, .
$$
In Fourier space, we define the vector $\hat{r}=Ur$, and analogously to Eq. \eqref{eq-Fourier-quad}, we have
\begin{equation}
\begin{split}
&\hat{r}_{2k-1}=\varphi_{q_k}:=\frac{1}{\sqrt{L}}\sum_{j=1}^L e^{{\rm i} q_k j} w^1_{j}\, , \\ &\hat{r}_{2k}=\pi_{q_k}:=\frac{1}{\sqrt{L}}\sum_{j=1}^L e^{{\rm i} q_k j} w^2_{j}\, ,
\label{eq-Fourier-quad-fer}
\end{split}
\end{equation}
for $k=1,2,3,\dots L$. Recall that $q_k=2\pi k /L$. The action of the Hamiltonian generator on these operators gives 
$$
{\rm i}[E,\hat{r}_i]=4\sum_{j=1}^{2L}\hat{H}_{ij}\hat{r}_j\, ,
$$ 
where $\hat{H}$  is block-diagonal with $2\times2$ blocks $\hat{h}_{q_k}$, 
$$
\hat{h}_{q_k}=h_0+h_1 e^{-iq_k}-h_1^T e^{{\rm i} q_k}=\frac{J}{2}\begin{pmatrix}
0&-h^x+ e^{{\rm i}q_k}\\
h^x- e^{-{\rm i} q_k}&0
\end{pmatrix} .
$$
To find the eigenmodes of the Hamiltonian we consider (for each $k$)
\begin{equation}
{\rm i}\left[E,\begin{pmatrix}
\varphi_{q_k}\\
\pi_{q_k}
\end{pmatrix}\right]=4\hat{h}_{q_k}\begin{pmatrix}
\varphi_{q_k}\\
\pi_{q_k}
\end{pmatrix}\, .
\label{Ising-sm-1}
\end{equation}
We define $s_{12}=2J(e^{{\rm i}q_k}-h)$, and $\beta_{q_k}$ as the linear combination $\beta_{q_k}=u_1 \varphi_{q_k}+iu_2 \pi_{q_k}$. We require $\{\beta_{q_k},\beta_{q_k}^\dagger\}=|u_1|^2+|u_2|^2=1/2$. Using the result in Eq.~\ref{Ising-sm-1} and the definition of $s_{12}$ we find
$$
{\rm i}[E,\beta_{q_k}]=u_1 s_{12}\pi_{q_k}-{\rm i}u_2 s^*_{12}\varphi_{q_k}\stackrel{!}{=}-{\rm i}{\rm e}_{q_k}\beta_{q_k}\, .
$$
The last equality is what needs to be imposed to find $\beta_{q_k}$ as an eigenmode of the Hamiltonian. The function ${\rm e }_{q_k}$ is the dispersion relation, which also needs to be determined. This gives two further equations
\begin{equation}
\begin{split}
{\rm e}_{q_k}u_1=u_2 s^*_{12}\,\,  , \qquad u_2 {\rm e}_{q_k}=u_1 s_{12}\, .
\end{split}
\label{system-Ising-sm}
\end{equation}
From the second equation we find $u_1=u_2 {\rm e}_{q_k}/s_{12}$ which, once  inserted in the first one, gives 
$$
{\rm e}_{q_k}=\sqrt{|s_{12}|^2}=2J\sqrt{(h^x)^2+1-2h^x\cos (q_k)}\, .
$$
This is the well-known dispersion relation for the Ising chain. Given that ${\rm e}_{q_k}=|s_{12}|$, we can write 
$$
s_{12}={\rm e}_{q_k}e^{{\rm i}\theta_k}\, ,
$$
and thus, from the second equation above, we find $u_2=u_1 e^{{\rm i}\theta_k}$. Since we need to have $|u_1^2|+|u_2^2|=1/2$ and since an overall phase is not important in $\beta_{q_k}$, we take $u_1=1/2$ and $u_2=e^{{\rm i}\theta_k}/2$, where, for the sake of clarity, we have that 
$$
e^{{\rm i}\theta_k}=\frac{s_{12}}{{\rm e}_{q_k}}=\frac{e^{{\rm i}q_k}-h^x}{\sqrt{(h^x)^2+1-2h^x\cos (q_k)}}\, .
$$
It can then be straightforwardly checked that ${\rm i}[E,\beta_{q_k}]=-{\rm i e}_{q_k} \beta_{q_k}$.

In the same spirit, we can find the eigenmodes starting from the Fourier operators $\varphi_{-q_k},\pi_{-q_k}$. This is simply done by defining the operators $\hat{r}=U^\dagger r$. Rotating the matrix $H$ into this representation, we find 
\begin{equation}
{\rm i}\left[E,\begin{pmatrix}
\varphi_{-q_k}\\
\pi_{-q_k}
\end{pmatrix}\right]=4\left(\hat{h}_{q_k}\right)^*\begin{pmatrix}
\varphi_{-q_k}\\
\pi_{-q_k}
\end{pmatrix}\, .
\label{Ising-q-sm-1}
\end{equation}
One can  obtain the eigenmodes $\beta_{-q_k}$, as before. We collect all these operators and their Hermitean conjugates together:
\begin{equation}
\begin{split}
&\beta_{q_k}=\frac{1}{2}\left(\varphi_{q_k}+{\rm i}\frac{s_{12}}{{{\rm e}_{q_k}}}\pi_{q_k}\right)\, ,\\ &\beta_{-q_k}^\dagger=\frac{1}{2}\left(\varphi_{q_k}-{\rm i}\frac{s_{12}}{{{\rm e}_{q_k}}}\pi_{q_k}\right)\, ,\\
&\beta_{-q_k}=\frac{1}{2}\left(\varphi_{-q_k}+{\rm i}\frac{s_{12}^*}{{{\rm e}_{q_k}}}\pi_{-q_k}\right)\, ,\\
&\beta_{q_k}^\dagger=\frac{1}{2}\left(\varphi_{-q_k}-{\rm i}\frac{s_{12}^*}{{{\rm e}_{q_k}}}\pi_{-q_k}\right)\, .
\end{split}
\label{eig-mod-Ising}
\end{equation}
Inverting these relations, we find the expression for the Fourier operators $\varphi_{q_k},\pi_{q_k}$ and their Hermitean conjugates $\varphi_{-q_k},\pi_{-q_k}$
\begin{equation}
\begin{split}
\varphi_{q_k}&=(\beta_{q_k}+\beta_{-q_k}^\dagger)\, ,\qquad \pi_{q_k}={\rm i}(\beta_{-q_k}^\dagger-\beta_{q_k})\frac{{\rm e}_{q_k}}{s_{12}}\\
\varphi_{-q_k}&=(\beta_{-q_k}+\beta_{q_k}^\dagger)\, ,\qquad \pi_{-q_k}={\rm i}(\beta_{q_k}^\dagger-\beta_{-q_k})\frac{{\rm e}_{q_k}}{s_{12}}\, .
\end{split}
\end{equation}

For fermionic systems, the covariance matrix is defined as $G_{mn}=\braket{[r_m,r_n]}/2=\braket{r_m r_n}-\delta_{mn}$. In Fourier space, this becomes $$
\hat{G}=(UGU^\dagger)_{k h}=\braket{\hat{r}_{q_k} \hat{r}_{-q_h}}-\delta_{k h}\, .
$$
We consider as initial state the ground state of the Ising Hamiltonian for $h^x=h^x_0$. The quantities ${\rm e}_{q_k}^0$, $s_{12}^0$ (notice that also $s_{12}$ depends on $q_k$ even if this is not written explicitly) are associated to the value $h_0^x$ of the transverse field. It can be checked that the covariance matrix in Fourier space, for the ground state of $E$, is such that  
$$
\hat{g}_{q_k}=\begin{pmatrix}
0&{\rm i}\frac{{\rm e}_{q_k}^0}{(s_{12}^0)^*}\\
-{\rm i}\frac{{\rm e}_{q_k}^0}{s_{12}^0}&0
\end{pmatrix}\, .
$$
The real space covariance matrix is obtained as $G=U^\dagger \hat{G} U$ and determines the initial system state. 

For the time-evolution, we consider a quenched Hamiltonian field $h^x\neq h^x_0$ and the presence of non-local decay. The eigenmodes are the ones in Eq. \eqref{eig-mod-Ising}, and the velocities of the quasi-particles are
$$
v_{q_k}=\frac{4J^2 h^x \sin(q_k)}{{\rm e}_{q_k}}\, .
$$
To enable prediction from our method, we finally need to compute the action of the Lindblad generator on the number operator for quasi-particles $n_{q_k}=\beta_{q_k}^\dagger \beta_{q_k}$. This provides the rate equation for the densities of quasi-particles. We find, for non-local decay characterized by the function $f_{ij}$, the relation 
$$
\mathcal{L}\left[n_{q_k}\right]=-4\gamma_{q_k}n_{q_k}+2\gamma_{q_k}+\gamma_{q_k}\frac{s_{12}+s_{12}^*}{{\rm e}_{q_k}}\, ,
$$
with $\gamma_{q_k}=\gamma^-\mathcal{S}_k[f^-]$.
The densities of the quasi-particle for the quenched Hamiltonian in the initial state can be computed from the initial covariance matrix in Fourier space, as 
$$
\braket{n_{q_k}}=\frac{1}{2}-\frac{1}{4}\left[\frac{{\rm e}_{q_k}^0}{{\rm e}_{q_k}}\frac{s_{12}}{s_{12}^0}+\frac{{\rm e}_{q_k}^0}{{\rm e}_{q_k}}\frac{s_{12}^*}{(s_{12}^0)^*}\right]\, .
$$

In Fig.~\ref{Fig3}(d), we show numerical results for the Ising chain considered here. As displayed, our formula in Eq.~\eqref{propag} provides a good description for the time-evolution of the mutual information in the system. 

\bibliography{Notes_BIBLIO}

\begin{thebibliography}{83}%
\makeatletter
\providecommand \@ifxundefined [1]{%
 \@ifx{#1\undefined}
}%
\providecommand \@ifnum [1]{%
 \ifnum #1\expandafter \@firstoftwo
 \else \expandafter \@secondoftwo
 \fi
}%
\providecommand \@ifx [1]{%
 \ifx #1\expandafter \@firstoftwo
 \else \expandafter \@secondoftwo
 \fi
}%
\providecommand \natexlab [1]{#1}%
\providecommand \enquote  [1]{``#1''}%
\providecommand \bibnamefont  [1]{#1}%
\providecommand \bibfnamefont [1]{#1}%
\providecommand \citenamefont [1]{#1}%
\providecommand \href@noop [0]{\@secondoftwo}%
\providecommand \href [0]{\begingroup \@sanitize@url \@href}%
\providecommand \@href[1]{\@@startlink{#1}\@@href}%
\providecommand \@@href[1]{\endgroup#1\@@endlink}%
\providecommand \@sanitize@url [0]{\catcode `\\12\catcode `\$12\catcode
  `\&12\catcode `\#12\catcode `\^12\catcode `\_12\catcode `\%12\relax}%
\providecommand \@@startlink[1]{}%
\providecommand \@@endlink[0]{}%
\providecommand \url  [0]{\begingroup\@sanitize@url \@url }%
\providecommand \@url [1]{\endgroup\@href {#1}{\urlprefix }}%
\providecommand \urlprefix  [0]{URL }%
\providecommand \Eprint [0]{\href }%
\providecommand \doibase [0]{https://doi.org/}%
\providecommand \selectlanguage [0]{\@gobble}%
\providecommand \bibinfo  [0]{\@secondoftwo}%
\providecommand \bibfield  [0]{\@secondoftwo}%
\providecommand \translation [1]{[#1]}%
\providecommand \BibitemOpen [0]{}%
\providecommand \bibitemStop [0]{}%
\providecommand \bibitemNoStop [0]{.\EOS\space}%
\providecommand \EOS [0]{\spacefactor3000\relax}%
\providecommand \BibitemShut  [1]{\csname bibitem#1\endcsname}%
\let\auto@bib@innerbib\@empty
\bibitem [{\citenamefont {Wehrl}(1978)}]{wehrl1978}%
  \BibitemOpen
  \bibfield  {author} {\bibinfo {author} {\bibfnamefont {A.}~\bibnamefont
  {Wehrl}},\ }\bibfield  {title} {\bibinfo {title} {General properties of
  entropy},\ }\href {https://doi.org/10.1103/RevModPhys.50.221} {\bibfield
  {journal} {\bibinfo  {journal} {Rev. Mod. Phys.}\ }\textbf {\bibinfo {volume}
  {50}},\ \bibinfo {pages} {221} (\bibinfo {year} {1978})}\BibitemShut
  {NoStop}%
\bibitem [{\citenamefont {Shannon}(1948)}]{shannon1948}%
  \BibitemOpen
  \bibfield  {author} {\bibinfo {author} {\bibfnamefont {C.~E.}\ \bibnamefont
  {Shannon}},\ }\bibfield  {title} {\bibinfo {title} {A mathematical theory of
  communication},\ }\href {https://doi.org/10.1002/j.1538-7305.1948.tb01338.x}
  {\bibfield  {journal} {\bibinfo  {journal} {The Bell System Technical
  Journal}\ }\textbf {\bibinfo {volume} {27}},\ \bibinfo {pages} {379}
  (\bibinfo {year} {1948})}\BibitemShut {NoStop}%
\bibitem [{\citenamefont {Cover}\ and\ \citenamefont
  {Thomas}(1991)}]{cover1991}%
  \BibitemOpen
  \bibfield  {author} {\bibinfo {author} {\bibfnamefont {T.~M.}\ \bibnamefont
  {Cover}}\ and\ \bibinfo {author} {\bibfnamefont {J.~A.}\ \bibnamefont
  {Thomas}},\ }\href@noop {} {\emph {\bibinfo {title} {Elements of information
  theory}}},\ Vol.~\bibinfo {volume} {3}\ (\bibinfo  {publisher} {John Wiley \&
  Sons (New York)},\ \bibinfo {year} {1991})\BibitemShut {NoStop}%
\bibitem [{\citenamefont {Mezard}\ and\ \citenamefont
  {Montanari}(2009)}]{mezard2009}%
  \BibitemOpen
  \bibfield  {author} {\bibinfo {author} {\bibfnamefont {M.}~\bibnamefont
  {Mezard}}\ and\ \bibinfo {author} {\bibfnamefont {A.}~\bibnamefont
  {Montanari}},\ }\href@noop {} {\emph {\bibinfo {title} {Information, physics,
  and computation}}}\ (\bibinfo  {publisher} {Oxford University Press},\
  \bibinfo {year} {2009})\BibitemShut {NoStop}%
\bibitem [{\citenamefont {Gray}(2011)}]{gray2011}%
  \BibitemOpen
  \bibfield  {author} {\bibinfo {author} {\bibfnamefont {R.~M.}\ \bibnamefont
  {Gray}},\ }\href@noop {} {\emph {\bibinfo {title} {Entropy and information
  theory}}}\ (\bibinfo  {publisher} {Springer Science \& Business Media},\
  \bibinfo {year} {2011})\BibitemShut {NoStop}%
\bibitem [{\citenamefont {Calabrese}\ and\ \citenamefont
  {Cardy}(2005)}]{calabrese2005}%
  \BibitemOpen
  \bibfield  {author} {\bibinfo {author} {\bibfnamefont {P.}~\bibnamefont
  {Calabrese}}\ and\ \bibinfo {author} {\bibfnamefont {J.}~\bibnamefont
  {Cardy}},\ }\bibfield  {title} {\bibinfo {title} {Evolution of entanglement
  entropy in one-dimensional systems},\ }\href
  {https://doi.org/10.1088/1742-5468/2005/04/p04010} {\bibfield  {journal}
  {\bibinfo  {journal} {Journal of Statistical Mechanics: Theory and
  Experiment}\ }\textbf {\bibinfo {volume} {2005}},\ \bibinfo {pages} {P04010}
  (\bibinfo {year} {2005})}\BibitemShut {NoStop}%
\bibitem [{\citenamefont {Chiara}\ \emph {et~al.}(2006)\citenamefont {Chiara},
  \citenamefont {Montangero}, \citenamefont {Calabrese},\ and\ \citenamefont
  {Fazio}}]{chiara2006}%
  \BibitemOpen
  \bibfield  {author} {\bibinfo {author} {\bibfnamefont {G.~D.}\ \bibnamefont
  {Chiara}}, \bibinfo {author} {\bibfnamefont {S.}~\bibnamefont {Montangero}},
  \bibinfo {author} {\bibfnamefont {P.}~\bibnamefont {Calabrese}},\ and\
  \bibinfo {author} {\bibfnamefont {R.}~\bibnamefont {Fazio}},\ }\bibfield
  {title} {\bibinfo {title} {Entanglement entropy dynamics of heisenberg
  chains},\ }\href {https://doi.org/10.1088/1742-5468/2006/03/p03001}
  {\bibfield  {journal} {\bibinfo  {journal} {Journal of Statistical Mechanics:
  Theory and Experiment}\ }\textbf {\bibinfo {volume} {2006}},\ \bibinfo
  {pages} {P03001} (\bibinfo {year} {2006})}\BibitemShut {NoStop}%
\bibitem [{\citenamefont {Calabrese}\ and\ \citenamefont
  {Cardy}(2007)}]{calabrese2007}%
  \BibitemOpen
  \bibfield  {author} {\bibinfo {author} {\bibfnamefont {P.}~\bibnamefont
  {Calabrese}}\ and\ \bibinfo {author} {\bibfnamefont {J.}~\bibnamefont
  {Cardy}},\ }\bibfield  {title} {\bibinfo {title} {Entanglement and
  correlation functions following a local quench: a conformal field theory
  approach},\ }\href {https://doi.org/10.1088/1742-5468/2007/10/p10004}
  {\bibfield  {journal} {\bibinfo  {journal} {Journal of Statistical Mechanics:
  Theory and Experiment}\ }\textbf {\bibinfo {volume} {2007}},\ \bibinfo
  {pages} {P10004} (\bibinfo {year} {2007})}\BibitemShut {NoStop}%
\bibitem [{\citenamefont {L{\"a}uchli}\ and\ \citenamefont
  {Kollath}(2008)}]{lauchli2008}%
  \BibitemOpen
  \bibfield  {author} {\bibinfo {author} {\bibfnamefont {A.~M.}\ \bibnamefont
  {L{\"a}uchli}}\ and\ \bibinfo {author} {\bibfnamefont {C.}~\bibnamefont
  {Kollath}},\ }\bibfield  {title} {\bibinfo {title} {Spreading of correlations
  and entanglement after a quench in the one-dimensional
  bose{\textendash}hubbard model},\ }\href
  {https://doi.org/10.1088/1742-5468/2008/05/p05018} {\bibfield  {journal}
  {\bibinfo  {journal} {Journal of Statistical Mechanics: Theory and
  Experiment}\ }\textbf {\bibinfo {volume} {2008}},\ \bibinfo {pages} {P05018}
  (\bibinfo {year} {2008})}\BibitemShut {NoStop}%
\bibitem [{\citenamefont {Manmana}\ \emph {et~al.}(2009)\citenamefont
  {Manmana}, \citenamefont {Wessel}, \citenamefont {Noack},\ and\ \citenamefont
  {Muramatsu}}]{manmana2009}%
  \BibitemOpen
  \bibfield  {author} {\bibinfo {author} {\bibfnamefont {S.~R.}\ \bibnamefont
  {Manmana}}, \bibinfo {author} {\bibfnamefont {S.}~\bibnamefont {Wessel}},
  \bibinfo {author} {\bibfnamefont {R.~M.}\ \bibnamefont {Noack}},\ and\
  \bibinfo {author} {\bibfnamefont {A.}~\bibnamefont {Muramatsu}},\ }\bibfield
  {title} {\bibinfo {title} {Time evolution of correlations in strongly
  interacting fermions after a quantum quench},\ }\href
  {https://doi.org/10.1103/PhysRevB.79.155104} {\bibfield  {journal} {\bibinfo
  {journal} {Phys. Rev. B}\ }\textbf {\bibinfo {volume} {79}},\ \bibinfo
  {pages} {155104} (\bibinfo {year} {2009})}\BibitemShut {NoStop}%
\bibitem [{\citenamefont {Cheneau}\ \emph {et~al.}(2012)\citenamefont
  {Cheneau}, \citenamefont {Barmettler}, \citenamefont {Poletti}, \citenamefont
  {Endres}, \citenamefont {Schau{\ss}}, \citenamefont {Fukuhara}, \citenamefont
  {Gross}, \citenamefont {Bloch}, \citenamefont {Kollath},\ and\ \citenamefont
  {Kuhr}}]{cheneau2012}%
  \BibitemOpen
  \bibfield  {author} {\bibinfo {author} {\bibfnamefont {M.}~\bibnamefont
  {Cheneau}}, \bibinfo {author} {\bibfnamefont {P.}~\bibnamefont {Barmettler}},
  \bibinfo {author} {\bibfnamefont {D.}~\bibnamefont {Poletti}}, \bibinfo
  {author} {\bibfnamefont {M.}~\bibnamefont {Endres}}, \bibinfo {author}
  {\bibfnamefont {P.}~\bibnamefont {Schau{\ss}}}, \bibinfo {author}
  {\bibfnamefont {T.}~\bibnamefont {Fukuhara}}, \bibinfo {author}
  {\bibfnamefont {C.}~\bibnamefont {Gross}}, \bibinfo {author} {\bibfnamefont
  {I.}~\bibnamefont {Bloch}}, \bibinfo {author} {\bibfnamefont
  {C.}~\bibnamefont {Kollath}},\ and\ \bibinfo {author} {\bibfnamefont
  {S.}~\bibnamefont {Kuhr}},\ }\bibfield  {title} {\bibinfo {title}
  {Light-cone-like spreading of correlations in a quantum many-body system},\
  }\href {https://doi.org/10.1038/nature10748} {\bibfield  {journal} {\bibinfo
  {journal} {Nature}\ }\textbf {\bibinfo {volume} {481}},\ \bibinfo {pages}
  {484} (\bibinfo {year} {2012})}\BibitemShut {NoStop}%
\bibitem [{\citenamefont {Calabrese}\ \emph
  {et~al.}(2012{\natexlab{a}})\citenamefont {Calabrese}, \citenamefont
  {Essler},\ and\ \citenamefont {Fagotti}}]{calabrese2012}%
  \BibitemOpen
  \bibfield  {author} {\bibinfo {author} {\bibfnamefont {P.}~\bibnamefont
  {Calabrese}}, \bibinfo {author} {\bibfnamefont {F.~H.~L.}\ \bibnamefont
  {Essler}},\ and\ \bibinfo {author} {\bibfnamefont {M.}~\bibnamefont
  {Fagotti}},\ }\bibfield  {title} {\bibinfo {title} {Quantum quench in the
  transverse field ising chain: I. time evolution of order parameter
  correlators},\ }\href {https://doi.org/10.1088/1742-5468/2012/07/p07016}
  {\bibfield  {journal} {\bibinfo  {journal} {Journal of Statistical Mechanics:
  Theory and Experiment}\ }\textbf {\bibinfo {volume} {2012}},\ \bibinfo
  {pages} {P07016} (\bibinfo {year} {2012}{\natexlab{a}})}\BibitemShut
  {NoStop}%
\bibitem [{\citenamefont {Calabrese}\ \emph
  {et~al.}(2012{\natexlab{b}})\citenamefont {Calabrese}, \citenamefont
  {Essler},\ and\ \citenamefont {Fagotti}}]{calabrese2012b}%
  \BibitemOpen
  \bibfield  {author} {\bibinfo {author} {\bibfnamefont {P.}~\bibnamefont
  {Calabrese}}, \bibinfo {author} {\bibfnamefont {F.~H.~L.}\ \bibnamefont
  {Essler}},\ and\ \bibinfo {author} {\bibfnamefont {M.}~\bibnamefont
  {Fagotti}},\ }\bibfield  {title} {\bibinfo {title} {Quantum quenches in the
  transverse field ising chain: Ii. stationary state properties},\ }\href
  {https://doi.org/10.1088/1742-5468/2012/07/p07022} {\bibfield  {journal}
  {\bibinfo  {journal} {Journal of Statistical Mechanics: Theory and
  Experiment}\ }\textbf {\bibinfo {volume} {2012}},\ \bibinfo {pages} {P07022}
  (\bibinfo {year} {2012}{\natexlab{b}})}\BibitemShut {NoStop}%
\bibitem [{\citenamefont {Jurcevic}\ \emph {et~al.}(2014)\citenamefont
  {Jurcevic}, \citenamefont {Lanyon}, \citenamefont {Hauke}, \citenamefont
  {Hempel}, \citenamefont {Zoller}, \citenamefont {Blatt},\ and\ \citenamefont
  {Roos}}]{jurcevic2014}%
  \BibitemOpen
  \bibfield  {author} {\bibinfo {author} {\bibfnamefont {P.}~\bibnamefont
  {Jurcevic}}, \bibinfo {author} {\bibfnamefont {B.~P.}\ \bibnamefont
  {Lanyon}}, \bibinfo {author} {\bibfnamefont {P.}~\bibnamefont {Hauke}},
  \bibinfo {author} {\bibfnamefont {C.}~\bibnamefont {Hempel}}, \bibinfo
  {author} {\bibfnamefont {P.}~\bibnamefont {Zoller}}, \bibinfo {author}
  {\bibfnamefont {R.}~\bibnamefont {Blatt}},\ and\ \bibinfo {author}
  {\bibfnamefont {C.~F.}\ \bibnamefont {Roos}},\ }\bibfield  {title} {\bibinfo
  {title} {Quasiparticle engineering and entanglement propagation in a quantum
  many-body system},\ }\href {https://doi.org/10.1038/nature13461} {\bibfield
  {journal} {\bibinfo  {journal} {Nature}\ }\textbf {\bibinfo {volume} {511}},\
  \bibinfo {pages} {202} (\bibinfo {year} {2014})}\BibitemShut {NoStop}%
\bibitem [{\citenamefont {Richerme}\ \emph {et~al.}(2014)\citenamefont
  {Richerme}, \citenamefont {Gong}, \citenamefont {Lee}, \citenamefont {Senko},
  \citenamefont {Smith}, \citenamefont {Foss-Feig}, \citenamefont {Michalakis},
  \citenamefont {Gorshkov},\ and\ \citenamefont {Monroe}}]{richerme2014}%
  \BibitemOpen
  \bibfield  {author} {\bibinfo {author} {\bibfnamefont {P.}~\bibnamefont
  {Richerme}}, \bibinfo {author} {\bibfnamefont {Z.-X.}\ \bibnamefont {Gong}},
  \bibinfo {author} {\bibfnamefont {A.}~\bibnamefont {Lee}}, \bibinfo {author}
  {\bibfnamefont {C.}~\bibnamefont {Senko}}, \bibinfo {author} {\bibfnamefont
  {J.}~\bibnamefont {Smith}}, \bibinfo {author} {\bibfnamefont
  {M.}~\bibnamefont {Foss-Feig}}, \bibinfo {author} {\bibfnamefont
  {S.}~\bibnamefont {Michalakis}}, \bibinfo {author} {\bibfnamefont {A.~V.}\
  \bibnamefont {Gorshkov}},\ and\ \bibinfo {author} {\bibfnamefont
  {C.}~\bibnamefont {Monroe}},\ }\bibfield  {title} {\bibinfo {title}
  {Non-local propagation of correlations in quantum systems with long-range
  interactions},\ }\href {https://doi.org/10.1038/nature13450} {\bibfield
  {journal} {\bibinfo  {journal} {Nature}\ }\textbf {\bibinfo {volume} {511}},\
  \bibinfo {pages} {198} (\bibinfo {year} {2014})}\BibitemShut {NoStop}%
\bibitem [{\citenamefont {Islam}\ \emph {et~al.}(2015)\citenamefont {Islam},
  \citenamefont {Ma}, \citenamefont {Preiss}, \citenamefont {Eric~Tai},
  \citenamefont {Lukin}, \citenamefont {Rispoli},\ and\ \citenamefont
  {Greiner}}]{islam2015}%
  \BibitemOpen
  \bibfield  {author} {\bibinfo {author} {\bibfnamefont {R.}~\bibnamefont
  {Islam}}, \bibinfo {author} {\bibfnamefont {R.}~\bibnamefont {Ma}}, \bibinfo
  {author} {\bibfnamefont {P.~M.}\ \bibnamefont {Preiss}}, \bibinfo {author}
  {\bibfnamefont {M.}~\bibnamefont {Eric~Tai}}, \bibinfo {author}
  {\bibfnamefont {A.}~\bibnamefont {Lukin}}, \bibinfo {author} {\bibfnamefont
  {M.}~\bibnamefont {Rispoli}},\ and\ \bibinfo {author} {\bibfnamefont
  {M.}~\bibnamefont {Greiner}},\ }\bibfield  {title} {\bibinfo {title}
  {Measuring entanglement entropy in a quantum many-body system},\ }\href
  {https://doi.org/10.1038/nature15750} {\bibfield  {journal} {\bibinfo
  {journal} {Nature}\ }\textbf {\bibinfo {volume} {528}},\ \bibinfo {pages}
  {77} (\bibinfo {year} {2015})}\BibitemShut {NoStop}%
\bibitem [{\citenamefont {Eisert}\ \emph {et~al.}(2015)\citenamefont {Eisert},
  \citenamefont {Friesdorf},\ and\ \citenamefont {Gogolin}}]{eisert2015}%
  \BibitemOpen
  \bibfield  {author} {\bibinfo {author} {\bibfnamefont {J.}~\bibnamefont
  {Eisert}}, \bibinfo {author} {\bibfnamefont {M.}~\bibnamefont {Friesdorf}},\
  and\ \bibinfo {author} {\bibfnamefont {C.}~\bibnamefont {Gogolin}},\
  }\bibfield  {title} {\bibinfo {title} {Quantum many-body systems out of
  equilibrium},\ }\href {https://doi.org/10.1038/nphys3215} {\bibfield
  {journal} {\bibinfo  {journal} {Nature Physics}\ }\textbf {\bibinfo {volume}
  {11}},\ \bibinfo {pages} {124} (\bibinfo {year} {2015})}\BibitemShut
  {NoStop}%
\bibitem [{\citenamefont {Kaufman}\ \emph {et~al.}(2016)\citenamefont
  {Kaufman}, \citenamefont {Tai}, \citenamefont {Lukin}, \citenamefont
  {Rispoli}, \citenamefont {Schittko}, \citenamefont {Preiss},\ and\
  \citenamefont {Greiner}}]{kaufman2016}%
  \BibitemOpen
  \bibfield  {author} {\bibinfo {author} {\bibfnamefont {A.~M.}\ \bibnamefont
  {Kaufman}}, \bibinfo {author} {\bibfnamefont {M.~E.}\ \bibnamefont {Tai}},
  \bibinfo {author} {\bibfnamefont {A.}~\bibnamefont {Lukin}}, \bibinfo
  {author} {\bibfnamefont {M.}~\bibnamefont {Rispoli}}, \bibinfo {author}
  {\bibfnamefont {R.}~\bibnamefont {Schittko}}, \bibinfo {author}
  {\bibfnamefont {P.~M.}\ \bibnamefont {Preiss}},\ and\ \bibinfo {author}
  {\bibfnamefont {M.}~\bibnamefont {Greiner}},\ }\bibfield  {title} {\bibinfo
  {title} {Quantum thermalization through entanglement in an isolated many-body
  system},\ }\href {https://doi.org/10.1126/science.aaf6725} {\bibfield
  {journal} {\bibinfo  {journal} {Science}\ }\textbf {\bibinfo {volume}
  {353}},\ \bibinfo {pages} {794} (\bibinfo {year} {2016})}\BibitemShut
  {NoStop}%
\bibitem [{\citenamefont {Chang}\ \emph {et~al.}(2019)\citenamefont {Chang},
  \citenamefont {Chen}, \citenamefont {Gopalakrishnan},\ and\ \citenamefont
  {Pixley}}]{chang2019}%
  \BibitemOpen
  \bibfield  {author} {\bibinfo {author} {\bibfnamefont {P.-Y.}\ \bibnamefont
  {Chang}}, \bibinfo {author} {\bibfnamefont {X.}~\bibnamefont {Chen}},
  \bibinfo {author} {\bibfnamefont {S.}~\bibnamefont {Gopalakrishnan}},\ and\
  \bibinfo {author} {\bibfnamefont {J.~H.}\ \bibnamefont {Pixley}},\ }\bibfield
   {title} {\bibinfo {title} {Evolution of entanglement spectra under generic
  quantum dynamics},\ }\href {https://doi.org/10.1103/PhysRevLett.123.190602}
  {\bibfield  {journal} {\bibinfo  {journal} {Phys. Rev. Lett.}\ }\textbf
  {\bibinfo {volume} {123}},\ \bibinfo {pages} {190602} (\bibinfo {year}
  {2019})}\BibitemShut {NoStop}%
\bibitem [{\citenamefont {Rakovszky}\ \emph {et~al.}(2019)\citenamefont
  {Rakovszky}, \citenamefont {Pollmann},\ and\ \citenamefont {von
  Keyserlingk}}]{rakovszky2019}%
  \BibitemOpen
  \bibfield  {author} {\bibinfo {author} {\bibfnamefont {T.}~\bibnamefont
  {Rakovszky}}, \bibinfo {author} {\bibfnamefont {F.}~\bibnamefont
  {Pollmann}},\ and\ \bibinfo {author} {\bibfnamefont {C.~W.}\ \bibnamefont
  {von Keyserlingk}},\ }\bibfield  {title} {\bibinfo {title} {Sub-ballistic
  growth of r\'enyi entropies due to diffusion},\ }\href
  {https://doi.org/10.1103/PhysRevLett.122.250602} {\bibfield  {journal}
  {\bibinfo  {journal} {Phys. Rev. Lett.}\ }\textbf {\bibinfo {volume} {122}},\
  \bibinfo {pages} {250602} (\bibinfo {year} {2019})}\BibitemShut {NoStop}%
\bibitem [{\citenamefont {Brydges}\ \emph {et~al.}(2019)\citenamefont
  {Brydges}, \citenamefont {Elben}, \citenamefont {Jurcevic}, \citenamefont
  {Vermersch}, \citenamefont {Maier}, \citenamefont {Lanyon}, \citenamefont
  {Zoller}, \citenamefont {Blatt},\ and\ \citenamefont {Roos}}]{brydges2019}%
  \BibitemOpen
  \bibfield  {author} {\bibinfo {author} {\bibfnamefont {T.}~\bibnamefont
  {Brydges}}, \bibinfo {author} {\bibfnamefont {A.}~\bibnamefont {Elben}},
  \bibinfo {author} {\bibfnamefont {P.}~\bibnamefont {Jurcevic}}, \bibinfo
  {author} {\bibfnamefont {B.}~\bibnamefont {Vermersch}}, \bibinfo {author}
  {\bibfnamefont {C.}~\bibnamefont {Maier}}, \bibinfo {author} {\bibfnamefont
  {B.~P.}\ \bibnamefont {Lanyon}}, \bibinfo {author} {\bibfnamefont
  {P.}~\bibnamefont {Zoller}}, \bibinfo {author} {\bibfnamefont
  {R.}~\bibnamefont {Blatt}},\ and\ \bibinfo {author} {\bibfnamefont {C.~F.}\
  \bibnamefont {Roos}},\ }\bibfield  {title} {\bibinfo {title} {Probing
  r{\'e}nyi entanglement entropy via randomized measurements},\ }\href
  {https://doi.org/10.1126/science.aau4963} {\bibfield  {journal} {\bibinfo
  {journal} {Science}\ }\textbf {\bibinfo {volume} {364}},\ \bibinfo {pages}
  {260} (\bibinfo {year} {2019})}\BibitemShut {NoStop}%
\bibitem [{\citenamefont {Elben}\ \emph {et~al.}(2020)\citenamefont {Elben},
  \citenamefont {Kueng}, \citenamefont {Huang}, \citenamefont {van Bijnen},
  \citenamefont {Kokail}, \citenamefont {Dalmonte}, \citenamefont {Calabrese},
  \citenamefont {Kraus}, \citenamefont {Preskill}, \citenamefont {Zoller},\
  and\ \citenamefont {Vermersch}}]{elben2020}%
  \BibitemOpen
  \bibfield  {author} {\bibinfo {author} {\bibfnamefont {A.}~\bibnamefont
  {Elben}}, \bibinfo {author} {\bibfnamefont {R.}~\bibnamefont {Kueng}},
  \bibinfo {author} {\bibfnamefont {H.-Y.~R.}\ \bibnamefont {Huang}}, \bibinfo
  {author} {\bibfnamefont {R.}~\bibnamefont {van Bijnen}}, \bibinfo {author}
  {\bibfnamefont {C.}~\bibnamefont {Kokail}}, \bibinfo {author} {\bibfnamefont
  {M.}~\bibnamefont {Dalmonte}}, \bibinfo {author} {\bibfnamefont
  {P.}~\bibnamefont {Calabrese}}, \bibinfo {author} {\bibfnamefont
  {B.}~\bibnamefont {Kraus}}, \bibinfo {author} {\bibfnamefont
  {J.}~\bibnamefont {Preskill}}, \bibinfo {author} {\bibfnamefont
  {P.}~\bibnamefont {Zoller}},\ and\ \bibinfo {author} {\bibfnamefont
  {B.}~\bibnamefont {Vermersch}},\ }\bibfield  {title} {\bibinfo {title}
  {Mixed-state entanglement from local randomized measurements},\ }\href
  {https://doi.org/10.1103/PhysRevLett.125.200501} {\bibfield  {journal}
  {\bibinfo  {journal} {Phys. Rev. Lett.}\ }\textbf {\bibinfo {volume} {125}},\
  \bibinfo {pages} {200501} (\bibinfo {year} {2020})}\BibitemShut {NoStop}%
\bibitem [{\citenamefont {Gillman}\ \emph {et~al.}(2021)\citenamefont
  {Gillman}, \citenamefont {Carollo},\ and\ \citenamefont
  {Lesanovsky}}]{gillman2021}%
  \BibitemOpen
  \bibfield  {author} {\bibinfo {author} {\bibfnamefont {E.}~\bibnamefont
  {Gillman}}, \bibinfo {author} {\bibfnamefont {F.}~\bibnamefont {Carollo}},\
  and\ \bibinfo {author} {\bibfnamefont {I.}~\bibnamefont {Lesanovsky}},\
  }\bibfield  {title} {\bibinfo {title} {Quantum and classical temporal
  correlations in $(1 + 1)d$ quantum cellular automata},\ }\href@noop {}
  {\bibfield  {journal} {\bibinfo  {journal} {arXiv:2104.04279}\ } (\bibinfo
  {year} {2021})}\BibitemShut {NoStop}%
\bibitem [{\citenamefont {De~Nicola}\ \emph {et~al.}(2021)\citenamefont
  {De~Nicola}, \citenamefont {Michailidis},\ and\ \citenamefont
  {Serbyn}}]{denicola2021}%
  \BibitemOpen
  \bibfield  {author} {\bibinfo {author} {\bibfnamefont {S.}~\bibnamefont
  {De~Nicola}}, \bibinfo {author} {\bibfnamefont {A.~A.}\ \bibnamefont
  {Michailidis}},\ and\ \bibinfo {author} {\bibfnamefont {M.}~\bibnamefont
  {Serbyn}},\ }\bibfield  {title} {\bibinfo {title} {Entanglement view of
  dynamical quantum phase transitions},\ }\href
  {https://doi.org/10.1103/PhysRevLett.126.040602} {\bibfield  {journal}
  {\bibinfo  {journal} {Phys. Rev. Lett.}\ }\textbf {\bibinfo {volume} {126}},\
  \bibinfo {pages} {040602} (\bibinfo {year} {2021})}\BibitemShut {NoStop}%
\bibitem [{\citenamefont {Alba}\ \emph {et~al.}(2021)\citenamefont {Alba},
  \citenamefont {Bertini}, \citenamefont {Fagotti}, \citenamefont {Piroli},\
  and\ \citenamefont {Ruggiero}}]{alba2021b}%
  \BibitemOpen
  \bibfield  {author} {\bibinfo {author} {\bibfnamefont {V.}~\bibnamefont
  {Alba}}, \bibinfo {author} {\bibfnamefont {B.}~\bibnamefont {Bertini}},
  \bibinfo {author} {\bibfnamefont {M.}~\bibnamefont {Fagotti}}, \bibinfo
  {author} {\bibfnamefont {L.}~\bibnamefont {Piroli}},\ and\ \bibinfo {author}
  {\bibfnamefont {P.}~\bibnamefont {Ruggiero}},\ }\bibfield  {title} {\bibinfo
  {title} {Generalized-hydrodynamic approach to inhomogeneous quenches:
  Correlations, entanglement and quantum effects},\ }\href@noop {} {\bibfield
  {journal} {\bibinfo  {journal} {arXiv:2104.00656}\ } (\bibinfo {year}
  {2021})}\BibitemShut {NoStop}%
\bibitem [{\citenamefont {Mendoza-Arenas}\ and\ \citenamefont {Bu{\v
  c}a}(2021)}]{mendozaarenas2021}%
  \BibitemOpen
  \bibfield  {author} {\bibinfo {author} {\bibfnamefont {J.~J.}\ \bibnamefont
  {Mendoza-Arenas}}\ and\ \bibinfo {author} {\bibfnamefont {B.}~\bibnamefont
  {Bu{\v c}a}},\ }\bibfield  {title} {\bibinfo {title} {Self-induced
  entanglement resonance in a disordered bose-fermi mixture},\ }\href@noop {}
  {\bibfield  {journal} {\bibinfo  {journal} {arXiv:2106.06277}\ } (\bibinfo
  {year} {2021})}\BibitemShut {NoStop}%
\bibitem [{\citenamefont {Neel}\ \emph {et~al.}(2021)\citenamefont {Neel},
  \citenamefont {Yicheng}, \citenamefont {Yuan}, \citenamefont {Jerome},
  \citenamefont {Marcos},\ and\ \citenamefont {S.}}]{neel2021}%
  \BibitemOpen
  \bibfield  {author} {\bibinfo {author} {\bibfnamefont {M.}~\bibnamefont
  {Neel}}, \bibinfo {author} {\bibfnamefont {Z.}~\bibnamefont {Yicheng}},
  \bibinfo {author} {\bibfnamefont {L.}~\bibnamefont {Yuan}}, \bibinfo {author}
  {\bibfnamefont {D.}~\bibnamefont {Jerome}}, \bibinfo {author} {\bibfnamefont
  {R.}~\bibnamefont {Marcos}},\ and\ \bibinfo {author} {\bibfnamefont {W.~D.}\
  \bibnamefont {S.}},\ }\bibfield  {title} {\bibinfo {title} {Generalized
  hydrodynamics in strongly interacting 1d bose gases},\ }\href
  {https://doi.org/10.1126/science.abf0147} {\bibfield  {journal} {\bibinfo
  {journal} {Science}\ }\textbf {\bibinfo {volume} {373}},\ \bibinfo {pages}
  {1129} (\bibinfo {year} {2021})}\BibitemShut {NoStop}%
\bibitem [{\citenamefont {Bernier}\ \emph {et~al.}(2018)\citenamefont
  {Bernier}, \citenamefont {Tan}, \citenamefont {Bonnes}, \citenamefont {Guo},
  \citenamefont {Poletti},\ and\ \citenamefont {Kollath}}]{bernier2018}%
  \BibitemOpen
  \bibfield  {author} {\bibinfo {author} {\bibfnamefont {J.-S.}\ \bibnamefont
  {Bernier}}, \bibinfo {author} {\bibfnamefont {R.}~\bibnamefont {Tan}},
  \bibinfo {author} {\bibfnamefont {L.}~\bibnamefont {Bonnes}}, \bibinfo
  {author} {\bibfnamefont {C.}~\bibnamefont {Guo}}, \bibinfo {author}
  {\bibfnamefont {D.}~\bibnamefont {Poletti}},\ and\ \bibinfo {author}
  {\bibfnamefont {C.}~\bibnamefont {Kollath}},\ }\bibfield  {title} {\bibinfo
  {title} {Light-cone and diffusive propagation of correlations in a many-body
  dissipative system},\ }\href {https://doi.org/10.1103/PhysRevLett.120.020401}
  {\bibfield  {journal} {\bibinfo  {journal} {Phys. Rev. Lett.}\ }\textbf
  {\bibinfo {volume} {120}},\ \bibinfo {pages} {020401} (\bibinfo {year}
  {2018})}\BibitemShut {NoStop}%
\bibitem [{\citenamefont {Macieszczak}\ \emph {et~al.}(2019)\citenamefont
  {Macieszczak}, \citenamefont {Levi}, \citenamefont {Macr\`{\i}},
  \citenamefont {Lesanovsky},\ and\ \citenamefont
  {Garrahan}}]{macieszczak2019}%
  \BibitemOpen
  \bibfield  {author} {\bibinfo {author} {\bibfnamefont {K.}~\bibnamefont
  {Macieszczak}}, \bibinfo {author} {\bibfnamefont {E.}~\bibnamefont {Levi}},
  \bibinfo {author} {\bibfnamefont {T.}~\bibnamefont {Macr\`{\i}}}, \bibinfo
  {author} {\bibfnamefont {I.}~\bibnamefont {Lesanovsky}},\ and\ \bibinfo
  {author} {\bibfnamefont {J.~P.}\ \bibnamefont {Garrahan}},\ }\bibfield
  {title} {\bibinfo {title} {Coherence, entanglement, and quantumness in closed
  and open systems with conserved charge, with an application to many-body
  localization},\ }\href {https://doi.org/10.1103/PhysRevA.99.052354}
  {\bibfield  {journal} {\bibinfo  {journal} {Phys. Rev. A}\ }\textbf {\bibinfo
  {volume} {99}},\ \bibinfo {pages} {052354} (\bibinfo {year}
  {2019})}\BibitemShut {NoStop}%
\bibitem [{\citenamefont {Malouf}\ \emph {et~al.}(2020)\citenamefont {Malouf},
  \citenamefont {Goold}, \citenamefont {Adesso},\ and\ \citenamefont
  {Landi}}]{malouf2020}%
  \BibitemOpen
  \bibfield  {author} {\bibinfo {author} {\bibfnamefont {W.~T.~B.}\
  \bibnamefont {Malouf}}, \bibinfo {author} {\bibfnamefont {J.}~\bibnamefont
  {Goold}}, \bibinfo {author} {\bibfnamefont {G.}~\bibnamefont {Adesso}},\ and\
  \bibinfo {author} {\bibfnamefont {G.~T.}\ \bibnamefont {Landi}},\ }\bibfield
  {title} {\bibinfo {title} {Analysis of the conditional mutual information in
  ballistic and diffusive non-equilibrium steady-states},\ }\href
  {https://doi.org/10.1088/1751-8121/ab93fd} {\bibfield  {journal} {\bibinfo
  {journal} {Journal of Physics A: Mathematical and Theoretical}\ }\textbf
  {\bibinfo {volume} {53}},\ \bibinfo {pages} {305302} (\bibinfo {year}
  {2020})}\BibitemShut {NoStop}%
\bibitem [{\citenamefont {Maity}\ \emph {et~al.}(2020)\citenamefont {Maity},
  \citenamefont {Bandyopadhyay}, \citenamefont {Bhattacharjee},\ and\
  \citenamefont {Dutta}}]{maity2020}%
  \BibitemOpen
  \bibfield  {author} {\bibinfo {author} {\bibfnamefont {S.}~\bibnamefont
  {Maity}}, \bibinfo {author} {\bibfnamefont {S.}~\bibnamefont
  {Bandyopadhyay}}, \bibinfo {author} {\bibfnamefont {S.}~\bibnamefont
  {Bhattacharjee}},\ and\ \bibinfo {author} {\bibfnamefont {A.}~\bibnamefont
  {Dutta}},\ }\bibfield  {title} {\bibinfo {title} {Growth of mutual
  information in a quenched one-dimensional open quantum many-body system},\
  }\href {https://doi.org/10.1103/PhysRevB.101.180301} {\bibfield  {journal}
  {\bibinfo  {journal} {Phys. Rev. B}\ }\textbf {\bibinfo {volume} {101}},\
  \bibinfo {pages} {180301} (\bibinfo {year} {2020})}\BibitemShut {NoStop}%
\bibitem [{\citenamefont {Alba}\ and\ \citenamefont
  {Carollo}(2021{\natexlab{a}})}]{alba2021}%
  \BibitemOpen
  \bibfield  {author} {\bibinfo {author} {\bibfnamefont {V.}~\bibnamefont
  {Alba}}\ and\ \bibinfo {author} {\bibfnamefont {F.}~\bibnamefont {Carollo}},\
  }\bibfield  {title} {\bibinfo {title} {Spreading of correlations in markovian
  open quantum systems},\ }\href {https://doi.org/10.1103/PhysRevB.103.L020302}
  {\bibfield  {journal} {\bibinfo  {journal} {Phys. Rev. B}\ }\textbf {\bibinfo
  {volume} {103}},\ \bibinfo {pages} {L020302} (\bibinfo {year}
  {2021}{\natexlab{a}})}\BibitemShut {NoStop}%
\bibitem [{\citenamefont {Rossini}\ and\ \citenamefont
  {Vicari}(2021)}]{rossini2021}%
  \BibitemOpen
  \bibfield  {author} {\bibinfo {author} {\bibfnamefont {D.}~\bibnamefont
  {Rossini}}\ and\ \bibinfo {author} {\bibfnamefont {E.}~\bibnamefont
  {Vicari}},\ }\bibfield  {title} {\bibinfo {title} {Coherent and dissipative
  dynamics at quantum phase transitions},\ }\href@noop {} {\bibfield  {journal}
  {\bibinfo  {journal} {arXiv:2103.02626}\ } (\bibinfo {year}
  {2021})}\BibitemShut {NoStop}%
\bibitem [{\citenamefont {Li}\ \emph {et~al.}(2018)\citenamefont {Li},
  \citenamefont {Chen},\ and\ \citenamefont {Fisher}}]{li2018}%
  \BibitemOpen
  \bibfield  {author} {\bibinfo {author} {\bibfnamefont {Y.}~\bibnamefont
  {Li}}, \bibinfo {author} {\bibfnamefont {X.}~\bibnamefont {Chen}},\ and\
  \bibinfo {author} {\bibfnamefont {M.~P.~A.}\ \bibnamefont {Fisher}},\
  }\bibfield  {title} {\bibinfo {title} {Quantum zeno effect and the many-body
  entanglement transition},\ }\href
  {https://doi.org/10.1103/PhysRevB.98.205136} {\bibfield  {journal} {\bibinfo
  {journal} {Phys. Rev. B}\ }\textbf {\bibinfo {volume} {98}},\ \bibinfo
  {pages} {205136} (\bibinfo {year} {2018})}\BibitemShut {NoStop}%
\bibitem [{\citenamefont {{\v Z}nidari{\v c}}(2020)}]{znidaric2020}%
  \BibitemOpen
  \bibfield  {author} {\bibinfo {author} {\bibfnamefont {M.}~\bibnamefont {{\v
  Z}nidari{\v c}}},\ }\bibfield  {title} {\bibinfo {title} {Entanglement growth
  in diffusive systems},\ }\href {https://doi.org/10.1038/s42005-020-0366-7}
  {\bibfield  {journal} {\bibinfo  {journal} {Communications Physics}\ }\textbf
  {\bibinfo {volume} {3}},\ \bibinfo {pages} {100} (\bibinfo {year}
  {2020})}\BibitemShut {NoStop}%
\bibitem [{\citenamefont {Chan}\ \emph {et~al.}(2019)\citenamefont {Chan},
  \citenamefont {Nandkishore}, \citenamefont {Pretko},\ and\ \citenamefont
  {Smith}}]{chan2019}%
  \BibitemOpen
  \bibfield  {author} {\bibinfo {author} {\bibfnamefont {A.}~\bibnamefont
  {Chan}}, \bibinfo {author} {\bibfnamefont {R.~M.}\ \bibnamefont
  {Nandkishore}}, \bibinfo {author} {\bibfnamefont {M.}~\bibnamefont
  {Pretko}},\ and\ \bibinfo {author} {\bibfnamefont {G.}~\bibnamefont
  {Smith}},\ }\bibfield  {title} {\bibinfo {title} {Unitary-projective
  entanglement dynamics},\ }\href {https://doi.org/10.1103/PhysRevB.99.224307}
  {\bibfield  {journal} {\bibinfo  {journal} {Phys. Rev. B}\ }\textbf {\bibinfo
  {volume} {99}},\ \bibinfo {pages} {224307} (\bibinfo {year}
  {2019})}\BibitemShut {NoStop}%
\bibitem [{\citenamefont {Skinner}\ \emph {et~al.}(2019)\citenamefont
  {Skinner}, \citenamefont {Ruhman},\ and\ \citenamefont
  {Nahum}}]{skinner2019}%
  \BibitemOpen
  \bibfield  {author} {\bibinfo {author} {\bibfnamefont {B.}~\bibnamefont
  {Skinner}}, \bibinfo {author} {\bibfnamefont {J.}~\bibnamefont {Ruhman}},\
  and\ \bibinfo {author} {\bibfnamefont {A.}~\bibnamefont {Nahum}},\ }\bibfield
   {title} {\bibinfo {title} {Measurement-induced phase transitions in the
  dynamics of entanglement},\ }\href
  {https://doi.org/10.1103/PhysRevX.9.031009} {\bibfield  {journal} {\bibinfo
  {journal} {Phys. Rev. X}\ }\textbf {\bibinfo {volume} {9}},\ \bibinfo {pages}
  {031009} (\bibinfo {year} {2019})}\BibitemShut {NoStop}%
\bibitem [{\citenamefont {Cao}\ \emph {et~al.}(2019)\citenamefont {Cao},
  \citenamefont {Tilloy},\ and\ \citenamefont {Luca}}]{cao2019}%
  \BibitemOpen
  \bibfield  {author} {\bibinfo {author} {\bibfnamefont {X.}~\bibnamefont
  {Cao}}, \bibinfo {author} {\bibfnamefont {A.}~\bibnamefont {Tilloy}},\ and\
  \bibinfo {author} {\bibfnamefont {A.~D.}\ \bibnamefont {Luca}},\ }\bibfield
  {title} {\bibinfo {title} {{Entanglement in a fermion chain under continuous
  monitoring}},\ }\href {https://doi.org/10.21468/SciPostPhys.7.2.024}
  {\bibfield  {journal} {\bibinfo  {journal} {SciPost Phys.}\ }\textbf
  {\bibinfo {volume} {7}},\ \bibinfo {pages} {24} (\bibinfo {year}
  {2019})}\BibitemShut {NoStop}%
\bibitem [{\citenamefont {Jian}\ \emph {et~al.}(2020)\citenamefont {Jian},
  \citenamefont {You}, \citenamefont {Vasseur},\ and\ \citenamefont
  {Ludwig}}]{jian2020}%
  \BibitemOpen
  \bibfield  {author} {\bibinfo {author} {\bibfnamefont {C.-M.}\ \bibnamefont
  {Jian}}, \bibinfo {author} {\bibfnamefont {Y.-Z.}\ \bibnamefont {You}},
  \bibinfo {author} {\bibfnamefont {R.}~\bibnamefont {Vasseur}},\ and\ \bibinfo
  {author} {\bibfnamefont {A.~W.~W.}\ \bibnamefont {Ludwig}},\ }\bibfield
  {title} {\bibinfo {title} {Measurement-induced criticality in random quantum
  circuits},\ }\href {https://doi.org/10.1103/PhysRevB.101.104302} {\bibfield
  {journal} {\bibinfo  {journal} {Phys. Rev. B}\ }\textbf {\bibinfo {volume}
  {101}},\ \bibinfo {pages} {104302} (\bibinfo {year} {2020})}\BibitemShut
  {NoStop}%
\bibitem [{\citenamefont {Carollo}\ and\ \citenamefont
  {P\'erez-Espigares}(2020)}]{carollo2020}%
  \BibitemOpen
  \bibfield  {author} {\bibinfo {author} {\bibfnamefont {F.}~\bibnamefont
  {Carollo}}\ and\ \bibinfo {author} {\bibfnamefont {C.}~\bibnamefont
  {P\'erez-Espigares}},\ }\bibfield  {title} {\bibinfo {title} {Entanglement
  statistics in markovian open quantum systems: A matter of mutation and
  selection},\ }\href {https://doi.org/10.1103/PhysRevE.102.030104} {\bibfield
  {journal} {\bibinfo  {journal} {Phys. Rev. E}\ }\textbf {\bibinfo {volume}
  {102}},\ \bibinfo {pages} {030104} (\bibinfo {year} {2020})}\BibitemShut
  {NoStop}%
\bibitem [{\citenamefont {Piroli}\ \emph {et~al.}(2020)\citenamefont {Piroli},
  \citenamefont {S{\"u}nderhauf},\ and\ \citenamefont {Qi}}]{piroli2020}%
  \BibitemOpen
  \bibfield  {author} {\bibinfo {author} {\bibfnamefont {L.}~\bibnamefont
  {Piroli}}, \bibinfo {author} {\bibfnamefont {C.}~\bibnamefont
  {S{\"u}nderhauf}},\ and\ \bibinfo {author} {\bibfnamefont {X.-L.}\
  \bibnamefont {Qi}},\ }\bibfield  {title} {\bibinfo {title} {A random unitary
  circuit model for black hole evaporation},\ }\href
  {https://doi.org/10.1007/JHEP04(2020)063} {\bibfield  {journal} {\bibinfo
  {journal} {Journal of High Energy Physics}\ }\textbf {\bibinfo {volume}
  {2020}},\ \bibinfo {pages} {63} (\bibinfo {year} {2020})}\BibitemShut
  {NoStop}%
\bibitem [{\citenamefont {Ippoliti}\ \emph {et~al.}(2021)\citenamefont
  {Ippoliti}, \citenamefont {Gullans}, \citenamefont {Gopalakrishnan},
  \citenamefont {Huse},\ and\ \citenamefont {Khemani}}]{ippoliti2021}%
  \BibitemOpen
  \bibfield  {author} {\bibinfo {author} {\bibfnamefont {M.}~\bibnamefont
  {Ippoliti}}, \bibinfo {author} {\bibfnamefont {M.~J.}\ \bibnamefont
  {Gullans}}, \bibinfo {author} {\bibfnamefont {S.}~\bibnamefont
  {Gopalakrishnan}}, \bibinfo {author} {\bibfnamefont {D.~A.}\ \bibnamefont
  {Huse}},\ and\ \bibinfo {author} {\bibfnamefont {V.}~\bibnamefont
  {Khemani}},\ }\bibfield  {title} {\bibinfo {title} {Entanglement phase
  transitions in measurement-only dynamics},\ }\href
  {https://doi.org/10.1103/PhysRevX.11.011030} {\bibfield  {journal} {\bibinfo
  {journal} {Phys. Rev. X}\ }\textbf {\bibinfo {volume} {11}},\ \bibinfo
  {pages} {011030} (\bibinfo {year} {2021})}\BibitemShut {NoStop}%
\bibitem [{\citenamefont {Nahum}\ \emph {et~al.}(2021)\citenamefont {Nahum},
  \citenamefont {Roy}, \citenamefont {Skinner},\ and\ \citenamefont
  {Ruhman}}]{nahum2021}%
  \BibitemOpen
  \bibfield  {author} {\bibinfo {author} {\bibfnamefont {A.}~\bibnamefont
  {Nahum}}, \bibinfo {author} {\bibfnamefont {S.}~\bibnamefont {Roy}}, \bibinfo
  {author} {\bibfnamefont {B.}~\bibnamefont {Skinner}},\ and\ \bibinfo {author}
  {\bibfnamefont {J.}~\bibnamefont {Ruhman}},\ }\bibfield  {title} {\bibinfo
  {title} {Measurement and entanglement phase transitions in all-to-all quantum
  circuits, on quantum trees, and in landau-ginsburg theory},\ }\href
  {https://doi.org/10.1103/PRXQuantum.2.010352} {\bibfield  {journal} {\bibinfo
   {journal} {PRX Quantum}\ }\textbf {\bibinfo {volume} {2}},\ \bibinfo {pages}
  {010352} (\bibinfo {year} {2021})}\BibitemShut {NoStop}%
\bibitem [{\citenamefont {Alberton}\ \emph {et~al.}(2021)\citenamefont
  {Alberton}, \citenamefont {Buchhold},\ and\ \citenamefont
  {Diehl}}]{alberton2021}%
  \BibitemOpen
  \bibfield  {author} {\bibinfo {author} {\bibfnamefont {O.}~\bibnamefont
  {Alberton}}, \bibinfo {author} {\bibfnamefont {M.}~\bibnamefont {Buchhold}},\
  and\ \bibinfo {author} {\bibfnamefont {S.}~\bibnamefont {Diehl}},\ }\bibfield
   {title} {\bibinfo {title} {Entanglement transition in a monitored
  free-fermion chain: From extended criticality to area law},\ }\href
  {https://doi.org/10.1103/PhysRevLett.126.170602} {\bibfield  {journal}
  {\bibinfo  {journal} {Phys. Rev. Lett.}\ }\textbf {\bibinfo {volume} {126}},\
  \bibinfo {pages} {170602} (\bibinfo {year} {2021})}\BibitemShut {NoStop}%
\bibitem [{\citenamefont {Lavasani}\ \emph {et~al.}(2021)\citenamefont
  {Lavasani}, \citenamefont {Alavirad},\ and\ \citenamefont
  {Barkeshli}}]{lavasani2021}%
  \BibitemOpen
  \bibfield  {author} {\bibinfo {author} {\bibfnamefont {A.}~\bibnamefont
  {Lavasani}}, \bibinfo {author} {\bibfnamefont {Y.}~\bibnamefont {Alavirad}},\
  and\ \bibinfo {author} {\bibfnamefont {M.}~\bibnamefont {Barkeshli}},\
  }\bibfield  {title} {\bibinfo {title} {Measurement-induced topological
  entanglement transitions in symmetric random quantum circuits},\ }\href
  {https://doi.org/10.1038/s41567-020-01112-z} {\bibfield  {journal} {\bibinfo
  {journal} {Nature Physics}\ }\textbf {\bibinfo {volume} {17}},\ \bibinfo
  {pages} {342} (\bibinfo {year} {2021})}\BibitemShut {NoStop}%
\bibitem [{\citenamefont {Bernard}\ and\ \citenamefont
  {Piroli}(2021)}]{bernard2021}%
  \BibitemOpen
  \bibfield  {author} {\bibinfo {author} {\bibfnamefont {D.}~\bibnamefont
  {Bernard}}\ and\ \bibinfo {author} {\bibfnamefont {L.}~\bibnamefont
  {Piroli}},\ }\bibfield  {title} {\bibinfo {title} {Entanglement distribution
  in the quantum symmetric simple exclusion process},\ }\href@noop {}
  {\bibfield  {journal} {\bibinfo  {journal} {arXiv: 2102.04745}\ } (\bibinfo
  {year} {2021})}\BibitemShut {NoStop}%
\bibitem [{\citenamefont {Lindblad}(1976)}]{lindblad1976}%
  \BibitemOpen
  \bibfield  {author} {\bibinfo {author} {\bibfnamefont {G.}~\bibnamefont
  {Lindblad}},\ }\bibfield  {title} {\bibinfo {title} {On the generators of
  quantum dynamical semigroups},\ }\href
  {https://projecteuclid.org:443/euclid.cmp/1103899849} {\bibfield  {journal}
  {\bibinfo  {journal} {Comm. Math. Phys.}\ }\textbf {\bibinfo {volume} {48}},\
  \bibinfo {pages} {119} (\bibinfo {year} {1976})}\BibitemShut {NoStop}%
\bibitem [{\citenamefont {Fagotti}\ and\ \citenamefont
  {Calabrese}(2008)}]{fagotti2008}%
  \BibitemOpen
  \bibfield  {author} {\bibinfo {author} {\bibfnamefont {M.}~\bibnamefont
  {Fagotti}}\ and\ \bibinfo {author} {\bibfnamefont {P.}~\bibnamefont
  {Calabrese}},\ }\bibfield  {title} {\bibinfo {title} {Evolution of
  entanglement entropy following a quantum quench: Analytic results for the
  $xy$ chain in a transverse magnetic field},\ }\href
  {https://doi.org/10.1103/PhysRevA.78.010306} {\bibfield  {journal} {\bibinfo
  {journal} {Phys. Rev. A}\ }\textbf {\bibinfo {volume} {78}},\ \bibinfo
  {pages} {010306} (\bibinfo {year} {2008})}\BibitemShut {NoStop}%
\bibitem [{\citenamefont {Alba}\ and\ \citenamefont
  {Calabrese}(2017{\natexlab{a}})}]{alba2017}%
  \BibitemOpen
  \bibfield  {author} {\bibinfo {author} {\bibfnamefont {V.}~\bibnamefont
  {Alba}}\ and\ \bibinfo {author} {\bibfnamefont {P.}~\bibnamefont
  {Calabrese}},\ }\bibfield  {title} {\bibinfo {title} {Entanglement and
  thermodynamics after a quantum quench in integrable systems},\ }\href
  {https://doi.org/10.1073/pnas.1703516114} {\bibfield  {journal} {\bibinfo
  {journal} {Proceedings of the National Academy of Sciences}\ }\textbf
  {\bibinfo {volume} {114}},\ \bibinfo {pages} {7947} (\bibinfo {year}
  {2017}{\natexlab{a}})}\BibitemShut {NoStop}%
\bibitem [{\citenamefont {Alba}\ and\ \citenamefont
  {Calabrese}(2017{\natexlab{b}})}]{alba2017b}%
  \BibitemOpen
  \bibfield  {author} {\bibinfo {author} {\bibfnamefont {V.}~\bibnamefont
  {Alba}}\ and\ \bibinfo {author} {\bibfnamefont {P.}~\bibnamefont
  {Calabrese}},\ }\bibfield  {title} {\bibinfo {title} {Quench action and
  r\'enyi entropies in integrable systems},\ }\href
  {https://doi.org/10.1103/PhysRevB.96.115421} {\bibfield  {journal} {\bibinfo
  {journal} {Phys. Rev. B}\ }\textbf {\bibinfo {volume} {96}},\ \bibinfo
  {pages} {115421} (\bibinfo {year} {2017}{\natexlab{b}})}\BibitemShut
  {NoStop}%
\bibitem [{\citenamefont {Alba}\ and\ \citenamefont
  {Calabrese}(2018)}]{alba2018}%
  \BibitemOpen
  \bibfield  {author} {\bibinfo {author} {\bibfnamefont {V.}~\bibnamefont
  {Alba}}\ and\ \bibinfo {author} {\bibfnamefont {P.}~\bibnamefont
  {Calabrese}},\ }\bibfield  {title} {\bibinfo {title} {{Entanglement dynamics
  after quantum quenches in generic integrable systems}},\ }\href
  {https://doi.org/10.21468/SciPostPhys.4.3.017} {\bibfield  {journal}
  {\bibinfo  {journal} {SciPost Phys.}\ }\textbf {\bibinfo {volume} {4}},\
  \bibinfo {pages} {17} (\bibinfo {year} {2018})}\BibitemShut {NoStop}%
\bibitem [{\citenamefont {Alba}\ and\ \citenamefont
  {Calabrese}(2019)}]{alba2019b}%
  \BibitemOpen
  \bibfield  {author} {\bibinfo {author} {\bibfnamefont {V.}~\bibnamefont
  {Alba}}\ and\ \bibinfo {author} {\bibfnamefont {P.}~\bibnamefont
  {Calabrese}},\ }\bibfield  {title} {\bibinfo {title} {Quantum information
  dynamics in multipartite integrable systems},\ }\href
  {https://doi.org/10.1209/0295-5075/126/60001} {\bibfield  {journal} {\bibinfo
   {journal} {{EPL} (Europhysics Letters)}\ }\textbf {\bibinfo {volume}
  {126}},\ \bibinfo {pages} {60001} (\bibinfo {year} {2019})}\BibitemShut
  {NoStop}%
\bibitem [{\citenamefont {Castro-Alvaredo}\ \emph {et~al.}(2018)\citenamefont
  {Castro-Alvaredo}, \citenamefont {De~Fazio}, \citenamefont {Doyon},\ and\
  \citenamefont {Sz\'ecs\'enyi}}]{castro-alvaredo2018}%
  \BibitemOpen
  \bibfield  {author} {\bibinfo {author} {\bibfnamefont {O.~A.}\ \bibnamefont
  {Castro-Alvaredo}}, \bibinfo {author} {\bibfnamefont {C.}~\bibnamefont
  {De~Fazio}}, \bibinfo {author} {\bibfnamefont {B.}~\bibnamefont {Doyon}},\
  and\ \bibinfo {author} {\bibfnamefont {I.~M.}\ \bibnamefont
  {Sz\'ecs\'enyi}},\ }\bibfield  {title} {\bibinfo {title} {Entanglement
  content of quasiparticle excitations},\ }\href
  {https://doi.org/10.1103/PhysRevLett.121.170602} {\bibfield  {journal}
  {\bibinfo  {journal} {Phys. Rev. Lett.}\ }\textbf {\bibinfo {volume} {121}},\
  \bibinfo {pages} {170602} (\bibinfo {year} {2018})}\BibitemShut {NoStop}%
\bibitem [{\citenamefont {Bertini}\ \emph
  {et~al.}(2018{\natexlab{a}})\citenamefont {Bertini}, \citenamefont {Fagotti},
  \citenamefont {Piroli},\ and\ \citenamefont {Calabrese}}]{bertini2018}%
  \BibitemOpen
  \bibfield  {author} {\bibinfo {author} {\bibfnamefont {B.}~\bibnamefont
  {Bertini}}, \bibinfo {author} {\bibfnamefont {M.}~\bibnamefont {Fagotti}},
  \bibinfo {author} {\bibfnamefont {L.}~\bibnamefont {Piroli}},\ and\ \bibinfo
  {author} {\bibfnamefont {P.}~\bibnamefont {Calabrese}},\ }\bibfield  {title}
  {\bibinfo {title} {Entanglement evolution and generalised hydrodynamics:
  noninteracting systems},\ }\href {https://doi.org/10.1088/1751-8121/aad82e}
  {\bibfield  {journal} {\bibinfo  {journal} {Journal of Physics A:
  Mathematical and Theoretical}\ }\textbf {\bibinfo {volume} {51}},\ \bibinfo
  {pages} {39LT01} (\bibinfo {year} {2018}{\natexlab{a}})}\BibitemShut
  {NoStop}%
\bibitem [{\citenamefont {Bertini}\ \emph
  {et~al.}(2018{\natexlab{b}})\citenamefont {Bertini}, \citenamefont
  {Tartaglia},\ and\ \citenamefont {Calabrese}}]{bertini2018b}%
  \BibitemOpen
  \bibfield  {author} {\bibinfo {author} {\bibfnamefont {B.}~\bibnamefont
  {Bertini}}, \bibinfo {author} {\bibfnamefont {E.}~\bibnamefont {Tartaglia}},\
  and\ \bibinfo {author} {\bibfnamefont {P.}~\bibnamefont {Calabrese}},\
  }\bibfield  {title} {\bibinfo {title} {Entanglement and diagonal entropies
  after a quench with no pair structure},\ }\href
  {https://doi.org/10.1088/1742-5468/aac73f} {\bibfield  {journal} {\bibinfo
  {journal} {Journal of Statistical Mechanics: Theory and Experiment}\ }\textbf
  {\bibinfo {volume} {2018}},\ \bibinfo {pages} {063104} (\bibinfo {year}
  {2018}{\natexlab{b}})}\BibitemShut {NoStop}%
\bibitem [{\citenamefont {Bastianello}\ and\ \citenamefont
  {Calabrese}(2018)}]{bastianello2018}%
  \BibitemOpen
  \bibfield  {author} {\bibinfo {author} {\bibfnamefont {A.}~\bibnamefont
  {Bastianello}}\ and\ \bibinfo {author} {\bibfnamefont {P.}~\bibnamefont
  {Calabrese}},\ }\bibfield  {title} {\bibinfo {title} {{Spreading of
  entanglement and correlations after a quench with intertwined
  quasiparticles}},\ }\href {https://doi.org/10.21468/SciPostPhys.5.4.033}
  {\bibfield  {journal} {\bibinfo  {journal} {SciPost Phys.}\ }\textbf
  {\bibinfo {volume} {5}},\ \bibinfo {pages} {33} (\bibinfo {year}
  {2018})}\BibitemShut {NoStop}%
\bibitem [{\citenamefont {Alba}\ \emph {et~al.}(2019)\citenamefont {Alba},
  \citenamefont {Bertini},\ and\ \citenamefont {Fagotti}}]{alba2019}%
  \BibitemOpen
  \bibfield  {author} {\bibinfo {author} {\bibfnamefont {V.}~\bibnamefont
  {Alba}}, \bibinfo {author} {\bibfnamefont {B.}~\bibnamefont {Bertini}},\ and\
  \bibinfo {author} {\bibfnamefont {M.}~\bibnamefont {Fagotti}},\ }\bibfield
  {title} {\bibinfo {title} {{Entanglement evolution and generalised
  hydrodynamics: interacting integrable systems}},\ }\href
  {https://doi.org/10.21468/SciPostPhys.7.1.005} {\bibfield  {journal}
  {\bibinfo  {journal} {SciPost Phys.}\ }\textbf {\bibinfo {volume} {7}},\
  \bibinfo {pages} {5} (\bibinfo {year} {2019})}\BibitemShut {NoStop}%
\bibitem [{\citenamefont {Bastianello}\ and\ \citenamefont
  {Collura}(2020)}]{bastianello2020}%
  \BibitemOpen
  \bibfield  {author} {\bibinfo {author} {\bibfnamefont {A.}~\bibnamefont
  {Bastianello}}\ and\ \bibinfo {author} {\bibfnamefont {M.}~\bibnamefont
  {Collura}},\ }\bibfield  {title} {\bibinfo {title} {{Entanglement spreading
  and quasiparticle picture beyond the pair structure}},\ }\href
  {https://doi.org/10.21468/SciPostPhys.8.3.045} {\bibfield  {journal}
  {\bibinfo  {journal} {SciPost Phys.}\ }\textbf {\bibinfo {volume} {8}},\
  \bibinfo {pages} {45} (\bibinfo {year} {2020})}\BibitemShut {NoStop}%
\bibitem [{\citenamefont {Yang}\ and\ \citenamefont {Yang}(1969)}]{yang1969}%
  \BibitemOpen
  \bibfield  {author} {\bibinfo {author} {\bibfnamefont {C.~N.}\ \bibnamefont
  {Yang}}\ and\ \bibinfo {author} {\bibfnamefont {C.~P.}\ \bibnamefont
  {Yang}},\ }\bibfield  {title} {\bibinfo {title} {Thermodynamics of a
  one‐dimensional system of bosons with repulsive delta‐function
  interaction},\ }\href {https://doi.org/10.1063/1.1664947} {\bibfield
  {journal} {\bibinfo  {journal} {Journal of Mathematical Physics}\ }\textbf
  {\bibinfo {volume} {10}},\ \bibinfo {pages} {1115} (\bibinfo {year}
  {1969})}\BibitemShut {NoStop}%
\bibitem [{\citenamefont {Polkovnikov}\ \emph {et~al.}(2011)\citenamefont
  {Polkovnikov}, \citenamefont {Sengupta}, \citenamefont {Silva},\ and\
  \citenamefont {Vengalattore}}]{polkovnikov2011}%
  \BibitemOpen
  \bibfield  {author} {\bibinfo {author} {\bibfnamefont {A.}~\bibnamefont
  {Polkovnikov}}, \bibinfo {author} {\bibfnamefont {K.}~\bibnamefont
  {Sengupta}}, \bibinfo {author} {\bibfnamefont {A.}~\bibnamefont {Silva}},\
  and\ \bibinfo {author} {\bibfnamefont {M.}~\bibnamefont {Vengalattore}},\
  }\bibfield  {title} {\bibinfo {title} {Colloquium: Nonequilibrium dynamics of
  closed interacting quantum systems},\ }\href
  {https://doi.org/10.1103/RevModPhys.83.863} {\bibfield  {journal} {\bibinfo
  {journal} {Rev. Mod. Phys.}\ }\textbf {\bibinfo {volume} {83}},\ \bibinfo
  {pages} {863} (\bibinfo {year} {2011})}\BibitemShut {NoStop}%
\bibitem [{\citenamefont {Calabrese}\ \emph {et~al.}(2016)\citenamefont
  {Calabrese}, \citenamefont {Essler},\ and\ \citenamefont
  {Mussardo}}]{calabrese2016}%
  \BibitemOpen
  \bibfield  {author} {\bibinfo {author} {\bibfnamefont {P.}~\bibnamefont
  {Calabrese}}, \bibinfo {author} {\bibfnamefont {F.~H.~L.}\ \bibnamefont
  {Essler}},\ and\ \bibinfo {author} {\bibfnamefont {G.}~\bibnamefont
  {Mussardo}},\ }\bibfield  {title} {\bibinfo {title} {Introduction to `quantum
  integrability in out of equilibrium systems'},\ }\href
  {https://doi.org/10.1088/1742-5468/2016/06/064001} {\bibfield  {journal}
  {\bibinfo  {journal} {Journal of Statistical Mechanics: Theory and
  Experiment}\ }\textbf {\bibinfo {volume} {2016}},\ \bibinfo {pages} {064001}
  (\bibinfo {year} {2016})}\BibitemShut {NoStop}%
\bibitem [{\citenamefont {Essler}\ and\ \citenamefont
  {Fagotti}(2016)}]{essler2016}%
  \BibitemOpen
  \bibfield  {author} {\bibinfo {author} {\bibfnamefont {F.~H.~L.}\
  \bibnamefont {Essler}}\ and\ \bibinfo {author} {\bibfnamefont
  {M.}~\bibnamefont {Fagotti}},\ }\bibfield  {title} {\bibinfo {title} {Quench
  dynamics and relaxation in isolated integrable quantum spin chains},\ }\href
  {https://doi.org/10.1088/1742-5468/2016/06/064002} {\bibfield  {journal}
  {\bibinfo  {journal} {Journal of Statistical Mechanics: Theory and
  Experiment}\ }\textbf {\bibinfo {volume} {2016}},\ \bibinfo {pages} {064002}
  (\bibinfo {year} {2016})}\BibitemShut {NoStop}%
\bibitem [{\citenamefont {Vidmar}\ and\ \citenamefont
  {Rigol}(2016)}]{vidmar2016}%
  \BibitemOpen
  \bibfield  {author} {\bibinfo {author} {\bibfnamefont {L.}~\bibnamefont
  {Vidmar}}\ and\ \bibinfo {author} {\bibfnamefont {M.}~\bibnamefont {Rigol}},\
  }\bibfield  {title} {\bibinfo {title} {Generalized gibbs ensemble in
  integrable lattice models},\ }\href
  {https://doi.org/10.1088/1742-5468/2016/06/064007} {\bibfield  {journal}
  {\bibinfo  {journal} {Journal of Statistical Mechanics: Theory and
  Experiment}\ }\textbf {\bibinfo {volume} {2016}},\ \bibinfo {pages} {064007}
  (\bibinfo {year} {2016})}\BibitemShut {NoStop}%
\bibitem [{\citenamefont {Caux}\ and\ \citenamefont {Essler}(2013)}]{caux2013}%
  \BibitemOpen
  \bibfield  {author} {\bibinfo {author} {\bibfnamefont {J.-S.}\ \bibnamefont
  {Caux}}\ and\ \bibinfo {author} {\bibfnamefont {F.~H.~L.}\ \bibnamefont
  {Essler}},\ }\bibfield  {title} {\bibinfo {title} {Time evolution of local
  observables after quenching to an integrable model},\ }\href
  {https://doi.org/10.1103/PhysRevLett.110.257203} {\bibfield  {journal}
  {\bibinfo  {journal} {Phys. Rev. Lett.}\ }\textbf {\bibinfo {volume} {110}},\
  \bibinfo {pages} {257203} (\bibinfo {year} {2013})}\BibitemShut {NoStop}%
\bibitem [{\citenamefont {Caux}(2016)}]{caux2016}%
  \BibitemOpen
  \bibfield  {author} {\bibinfo {author} {\bibfnamefont {J.-S.}\ \bibnamefont
  {Caux}},\ }\bibfield  {title} {\bibinfo {title} {The quench action},\ }\href
  {https://doi.org/10.1088/1742-5468/2016/06/064006} {\bibfield  {journal}
  {\bibinfo  {journal} {Journal of Statistical Mechanics: Theory and
  Experiment}\ }\textbf {\bibinfo {volume} {2016}},\ \bibinfo {pages} {064006}
  (\bibinfo {year} {2016})}\BibitemShut {NoStop}%
\bibitem [{\citenamefont {Prosen}(2008)}]{prosen2008}%
  \BibitemOpen
  \bibfield  {author} {\bibinfo {author} {\bibfnamefont {T.}~\bibnamefont
  {Prosen}},\ }\bibfield  {title} {\bibinfo {title} {Third quantization: a
  general method to solve master equations for quadratic open fermi systems},\
  }\href {https://doi.org/10.1088/1367-2630/10/4/043026} {\bibfield  {journal}
  {\bibinfo  {journal} {New Journal of Physics}\ }\textbf {\bibinfo {volume}
  {10}},\ \bibinfo {pages} {043026} (\bibinfo {year} {2008})}\BibitemShut
  {NoStop}%
\bibitem [{\citenamefont {Prosen}(2010)}]{prosen2010b}%
  \BibitemOpen
  \bibfield  {author} {\bibinfo {author} {\bibfnamefont {T.}~\bibnamefont
  {Prosen}},\ }\bibfield  {title} {\bibinfo {title} {Spectral theorem for the
  lindblad equation for quadratic open fermionic systems},\ }\href
  {https://doi.org/10.1088/1742-5468/2010/07/p07020} {\bibfield  {journal}
  {\bibinfo  {journal} {Journal of Statistical Mechanics: Theory and
  Experiment}\ }\textbf {\bibinfo {volume} {2010}},\ \bibinfo {pages} {P07020}
  (\bibinfo {year} {2010})}\BibitemShut {NoStop}%
\bibitem [{\citenamefont {Prosen}\ and\ \citenamefont
  {Seligman}(2010)}]{prosen2010}%
  \BibitemOpen
  \bibfield  {author} {\bibinfo {author} {\bibfnamefont {T.}~\bibnamefont
  {Prosen}}\ and\ \bibinfo {author} {\bibfnamefont {T.~H.}\ \bibnamefont
  {Seligman}},\ }\bibfield  {title} {\bibinfo {title} {Quantization over boson
  operator spaces},\ }\href {https://doi.org/10.1088/1751-8113/43/39/392004}
  {\bibfield  {journal} {\bibinfo  {journal} {Journal of Physics A:
  Mathematical and Theoretical}\ }\textbf {\bibinfo {volume} {43}},\ \bibinfo
  {pages} {392004} (\bibinfo {year} {2010})}\BibitemShut {NoStop}%
\bibitem [{\citenamefont {Kos}\ and\ \citenamefont {Prosen}(2017)}]{kos2017}%
  \BibitemOpen
  \bibfield  {author} {\bibinfo {author} {\bibfnamefont {P.}~\bibnamefont
  {Kos}}\ and\ \bibinfo {author} {\bibfnamefont {T.}~\bibnamefont {Prosen}},\
  }\bibfield  {title} {\bibinfo {title} {Time-dependent correlation functions
  in open quadratic fermionic systems},\ }\href
  {https://doi.org/10.1088/1742-5468/aa9681} {\bibfield  {journal} {\bibinfo
  {journal} {Journal of Statistical Mechanics: Theory and Experiment}\ }\textbf
  {\bibinfo {volume} {2017}},\ \bibinfo {pages} {123103} (\bibinfo {year}
  {2017})}\BibitemShut {NoStop}%
\bibitem [{\citenamefont {Guo}\ and\ \citenamefont {Poletti}(2017)}]{guo2017}%
  \BibitemOpen
  \bibfield  {author} {\bibinfo {author} {\bibfnamefont {C.}~\bibnamefont
  {Guo}}\ and\ \bibinfo {author} {\bibfnamefont {D.}~\bibnamefont {Poletti}},\
  }\bibfield  {title} {\bibinfo {title} {Solutions for bosonic and fermionic
  dissipative quadratic open systems},\ }\href
  {https://doi.org/10.1103/PhysRevA.95.052107} {\bibfield  {journal} {\bibinfo
  {journal} {Phys. Rev. A}\ }\textbf {\bibinfo {volume} {95}},\ \bibinfo
  {pages} {052107} (\bibinfo {year} {2017})}\BibitemShut {NoStop}%
\bibitem [{\citenamefont {Demoen}\ \emph {et~al.}(1979)\citenamefont {Demoen},
  \citenamefont {Vanheuverzwijn},\ and\ \citenamefont {Verbeure}}]{demoen1979}%
  \BibitemOpen
  \bibfield  {author} {\bibinfo {author} {\bibfnamefont {B.}~\bibnamefont
  {Demoen}}, \bibinfo {author} {\bibfnamefont {P.}~\bibnamefont
  {Vanheuverzwijn}},\ and\ \bibinfo {author} {\bibfnamefont {A.}~\bibnamefont
  {Verbeure}},\ }\bibfield  {title} {\bibinfo {title} {Completely positive
  quasi-free maps of the ccr-algebra},\ }\href
  {https://doi.org/https://doi.org/10.1016/0034-4877(79)90049-1} {\bibfield
  {journal} {\bibinfo  {journal} {Reports on Mathematical Physics}\ }\textbf
  {\bibinfo {volume} {15}},\ \bibinfo {pages} {27} (\bibinfo {year}
  {1979})}\BibitemShut {NoStop}%
\bibitem [{\citenamefont {Holevo}\ and\ \citenamefont
  {Werner}(2001)}]{holevo2001}%
  \BibitemOpen
  \bibfield  {author} {\bibinfo {author} {\bibfnamefont {A.~S.}\ \bibnamefont
  {Holevo}}\ and\ \bibinfo {author} {\bibfnamefont {R.~F.}\ \bibnamefont
  {Werner}},\ }\bibfield  {title} {\bibinfo {title} {Evaluating capacities of
  bosonic gaussian channels},\ }\href
  {https://doi.org/10.1103/PhysRevA.63.032312} {\bibfield  {journal} {\bibinfo
  {journal} {Phys. Rev. A}\ }\textbf {\bibinfo {volume} {63}},\ \bibinfo
  {pages} {032312} (\bibinfo {year} {2001})}\BibitemShut {NoStop}%
\bibitem [{\citenamefont {Hellmich}(2010)}]{hellmich2010}%
  \BibitemOpen
  \bibfield  {author} {\bibinfo {author} {\bibfnamefont {M.}~\bibnamefont
  {Hellmich}},\ }\bibfield  {title} {\bibinfo {title} {Quasi-free semigroups on
  the ccr algebra},\ }\href
  {https://doi.org/https://doi.org/10.1016/S0034-4877(10)80031-X} {\bibfield
  {journal} {\bibinfo  {journal} {Reports on Mathematical Physics}\ }\textbf
  {\bibinfo {volume} {66}},\ \bibinfo {pages} {277} (\bibinfo {year}
  {2010})}\BibitemShut {NoStop}%
\bibitem [{\citenamefont {Heinosaari}\ \emph {et~al.}(2010)\citenamefont
  {Heinosaari}, \citenamefont {Holevo},\ and\ \citenamefont
  {Wolf}}]{heinosaari2010}%
  \BibitemOpen
  \bibfield  {author} {\bibinfo {author} {\bibfnamefont {T.}~\bibnamefont
  {Heinosaari}}, \bibinfo {author} {\bibfnamefont {A.~S.}\ \bibnamefont
  {Holevo}},\ and\ \bibinfo {author} {\bibfnamefont {M.~M.}\ \bibnamefont
  {Wolf}},\ }\bibfield  {title} {\bibinfo {title} {The semigroup structure of
  gaussian channels},\ }\href@noop {} {\bibfield  {journal} {\bibinfo
  {journal} {Quantum Info. Comput.}\ }\textbf {\bibinfo {volume} {10}},\
  \bibinfo {pages} {619} (\bibinfo {year} {2010})}\BibitemShut {NoStop}%
\bibitem [{\citenamefont {Weedbrook}\ \emph {et~al.}(2012)\citenamefont
  {Weedbrook}, \citenamefont {Pirandola}, \citenamefont {Garc\'{\i}a-Patr\'on},
  \citenamefont {Cerf}, \citenamefont {Ralph}, \citenamefont {Shapiro},\ and\
  \citenamefont {Lloyd}}]{weedbrook2012}%
  \BibitemOpen
  \bibfield  {author} {\bibinfo {author} {\bibfnamefont {C.}~\bibnamefont
  {Weedbrook}}, \bibinfo {author} {\bibfnamefont {S.}~\bibnamefont
  {Pirandola}}, \bibinfo {author} {\bibfnamefont {R.}~\bibnamefont
  {Garc\'{\i}a-Patr\'on}}, \bibinfo {author} {\bibfnamefont {N.~J.}\
  \bibnamefont {Cerf}}, \bibinfo {author} {\bibfnamefont {T.~C.}\ \bibnamefont
  {Ralph}}, \bibinfo {author} {\bibfnamefont {J.~H.}\ \bibnamefont {Shapiro}},\
  and\ \bibinfo {author} {\bibfnamefont {S.}~\bibnamefont {Lloyd}},\ }\bibfield
   {title} {\bibinfo {title} {Gaussian quantum information},\ }\href
  {https://doi.org/10.1103/RevModPhys.84.621} {\bibfield  {journal} {\bibinfo
  {journal} {Rev. Mod. Phys.}\ }\textbf {\bibinfo {volume} {84}},\ \bibinfo
  {pages} {621} (\bibinfo {year} {2012})}\BibitemShut {NoStop}%
\bibitem [{\citenamefont {Parthasarathy}(2015)}]{parthasarathy2015}%
  \BibitemOpen
  \bibfield  {author} {\bibinfo {author} {\bibfnamefont {K.~R.}\ \bibnamefont
  {Parthasarathy}},\ }\bibfield  {title} {\bibinfo {title} {Symplectic
  dilations, gaussian states and gaussian channels},\ }\href
  {https://doi.org/10.1007/s13226-015-0144-5} {\bibfield  {journal} {\bibinfo
  {journal} {Indian Journal of Pure and Applied Mathematics}\ }\textbf
  {\bibinfo {volume} {46}},\ \bibinfo {pages} {419} (\bibinfo {year}
  {2015})}\BibitemShut {NoStop}%
\bibitem [{\citenamefont {Gorini}\ \emph {et~al.}(1976)\citenamefont {Gorini},
  \citenamefont {Kossakowski},\ and\ \citenamefont {Sudarshan}}]{gorini1976}%
  \BibitemOpen
  \bibfield  {author} {\bibinfo {author} {\bibfnamefont {V.}~\bibnamefont
  {Gorini}}, \bibinfo {author} {\bibfnamefont {A.}~\bibnamefont
  {Kossakowski}},\ and\ \bibinfo {author} {\bibfnamefont {E.~C.~G.}\
  \bibnamefont {Sudarshan}},\ }\bibfield  {title} {\bibinfo {title} {Completely
  positive dynamical semigroups of {N-level} systems},\ }\href@noop {}
  {\bibfield  {journal} {\bibinfo  {journal} {Journal of Mathematical Physics}\
  }\textbf {\bibinfo {volume} {17}},\ \bibinfo {pages} {821} (\bibinfo {year}
  {1976})}\BibitemShut {NoStop}%
\bibitem [{\citenamefont {Adesso}\ and\ \citenamefont
  {Illuminati}(2007)}]{adesso2007}%
  \BibitemOpen
  \bibfield  {author} {\bibinfo {author} {\bibfnamefont {G.}~\bibnamefont
  {Adesso}}\ and\ \bibinfo {author} {\bibfnamefont {F.}~\bibnamefont
  {Illuminati}},\ }\bibfield  {title} {\bibinfo {title} {Entanglement in
  continuous-variable systems: recent advances and current perspectives},\
  }\href {https://doi.org/10.1088/1751-8113/40/28/s01} {\bibfield  {journal}
  {\bibinfo  {journal} {Journal of Physics A: Mathematical and Theoretical}\
  }\textbf {\bibinfo {volume} {40}},\ \bibinfo {pages} {7821} (\bibinfo {year}
  {2007})}\BibitemShut {NoStop}%
\bibitem [{\citenamefont {Adesso}\ \emph {et~al.}(2014)\citenamefont {Adesso},
  \citenamefont {Ragy},\ and\ \citenamefont {Lee}}]{adesso2014}%
  \BibitemOpen
  \bibfield  {author} {\bibinfo {author} {\bibfnamefont {G.}~\bibnamefont
  {Adesso}}, \bibinfo {author} {\bibfnamefont {S.}~\bibnamefont {Ragy}},\ and\
  \bibinfo {author} {\bibfnamefont {A.~R.}\ \bibnamefont {Lee}},\ }\bibfield
  {title} {\bibinfo {title} {Continuous variable quantum information: Gaussian
  states and beyond},\ }\href {https://doi.org/10.1142/S1230161214400010}
  {\bibfield  {journal} {\bibinfo  {journal} {Open Systems \& Information
  Dynamics}\ }\textbf {\bibinfo {volume} {21}},\ \bibinfo {pages} {1440001}
  (\bibinfo {year} {2014})}\BibitemShut {NoStop}%
\bibitem [{\citenamefont {Peschel}\ and\ \citenamefont
  {Eisler}(2009)}]{peschel2009}%
  \BibitemOpen
  \bibfield  {author} {\bibinfo {author} {\bibfnamefont {I.}~\bibnamefont
  {Peschel}}\ and\ \bibinfo {author} {\bibfnamefont {V.}~\bibnamefont
  {Eisler}},\ }\bibfield  {title} {\bibinfo {title} {Reduced density matrices
  and entanglement entropy in free lattice models},\ }\href
  {https://doi.org/10.1088/1751-8113/42/50/504003} {\bibfield  {journal}
  {\bibinfo  {journal} {Journal of Physics A: Mathematical and Theoretical}\
  }\textbf {\bibinfo {volume} {42}},\ \bibinfo {pages} {504003} (\bibinfo
  {year} {2009})}\BibitemShut {NoStop}%
\bibitem [{\citenamefont {Calabrese}(2018)}]{calabrese2018}%
  \BibitemOpen
  \bibfield  {author} {\bibinfo {author} {\bibfnamefont {P.}~\bibnamefont
  {Calabrese}},\ }\bibfield  {title} {\bibinfo {title} {Entanglement and
  thermodynamics in non-equilibrium isolated quantum systems},\ }\href
  {https://doi.org/https://doi.org/10.1016/j.physa.2017.10.011} {\bibfield
  {journal} {\bibinfo  {journal} {Physica A: Statistical Mechanics and its
  Applications}\ }\textbf {\bibinfo {volume} {504}},\ \bibinfo {pages} {31}
  (\bibinfo {year} {2018})},\ \bibinfo {note} {lecture Notes of the 14th
  International Summer School on Fundamental Problems in Statistical
  Physics}\BibitemShut {NoStop}%
\bibitem [{\citenamefont {Coser}\ \emph {et~al.}(2014)\citenamefont {Coser},
  \citenamefont {Tonni},\ and\ \citenamefont {Calabrese}}]{coser2014}%
  \BibitemOpen
  \bibfield  {author} {\bibinfo {author} {\bibfnamefont {A.}~\bibnamefont
  {Coser}}, \bibinfo {author} {\bibfnamefont {E.}~\bibnamefont {Tonni}},\ and\
  \bibinfo {author} {\bibfnamefont {P.}~\bibnamefont {Calabrese}},\ }\bibfield
  {title} {\bibinfo {title} {Entanglement negativity after a global quantum
  quench},\ }\href {https://doi.org/10.1088/1742-5468/2014/12/p12017}
  {\bibfield  {journal} {\bibinfo  {journal} {Journal of Statistical Mechanics:
  Theory and Experiment}\ }\textbf {\bibinfo {volume} {2014}},\ \bibinfo
  {pages} {P12017} (\bibinfo {year} {2014})}\BibitemShut {NoStop}%
\bibitem [{\citenamefont {Alba}\ and\ \citenamefont
  {Carollo}(2021{\natexlab{b}})}]{alba2021c}%
  \BibitemOpen
  \bibfield  {author} {\bibinfo {author} {\bibfnamefont {V.}~\bibnamefont
  {Alba}}\ and\ \bibinfo {author} {\bibfnamefont {F.}~\bibnamefont {Carollo}},\
  }\bibfield  {title} {\bibinfo {title} {Hydrodynamics of quantum entropies in
  ising chains with linear dissipation},\ }\href@noop {} {\bibfield  {journal}
  {\bibinfo  {journal} {arXiv: 2109.01836}\ } (\bibinfo {year}
  {2021}{\natexlab{b}})}\BibitemShut {NoStop}%
\end{thebibliography}%

\end{document}